\documentclass[iop]{emulateapj}
\usepackage{graphicx}
\usepackage{color}
\usepackage{epsfig}

\begin{document}
%
%
%
%%%%%%%%%%%%%%%%%%%%%%%%%%%%%%%%%%%%%%%%%%%%%%%%%%%%%%%%%%%%%%%%%%%%%%%%%%%%%%%%%%%%
%%%%%%%%%%%%%%%%%%%%%%%%%%%%%%%%%%%%%%%%%%%%%%%%%%%%%%%%%%%%%%%%%%%%%%%%%%%%%%%%%%%%
%%%%%%%%%%%%%%%%%%%%%%%%%%%%%%%%%%%%%%%%%%%%%%%%%%%%%%%%%%%%%%%%%%%%%%%%%%%%%%%%%%%%
%%%%%%%%%%%%%%%%%%%%% NOTES %%%%%%%%%%%%%%%%%%%%%%%%%%%%%%%%%%%%%%%%%%%%%%%%%%%%%%%%
%%%%%%%%%%%%%%%%%%%%%%%%%%%%%%%%%%%%%%%%%%%%%%%%%%%%%%%%%%%%%%%%%%%%%%%%%%%%%%%%%%%%
% I think that command definitions are not allowed by the journal.
% The figures are encouraged to have eps format in the journal.
% I prefer to have the figures placed where it should be and not at the end of the manuscript.

\title{Propagating Waves Transverse to the Magnetic Field in a Solar Prominence}

\author{B. Schmieder\altaffilmark{1}, T.A. Kucera\altaffilmark{2}, K. Knizhnik \altaffilmark{3,2}, M. Luna \altaffilmark{4,5}, A. Lopez-Ariste \altaffilmark{6}\and D.Toot\altaffilmark{7}}

%\email{brigitte.schmieder@obspm.fr}

\altaffiltext{1}{Observatoire de Paris, LESIA, UMR 8109 (CNRS), 92195 Meudon, France}
\altaffiltext{2}{Code 671, NASA's GSFC, Greenbelt, MD, 20771 USA}
\altaffiltext{3}{Johns Hopkins University, Baltimore, MD USA}
\altaffiltext{4}{Instituto de Astrof\'{\i}sica de Canarias, E-38200 La Laguna, Tenerife, Spain}
\altaffiltext{5}{Universidad de La Laguna, Dept. Astrof\'{\i}sica, E-38206 La Laguna, Tenerife, Spain}
\altaffiltext{6}{THEMIS, CNRS-UPS853, E38205 LaLaguna, Spain}
\altaffiltext{7}{Alfred University, Alfred, NY, USA}

\begin{abstract}
We report an unusual set of observations of waves in a large prominence
pillar which consist of pulses propagating perpendicular to the prominence
magnetic field. We  observe a huge   quiescent prominence with  the Solar Dynamics Observatory (SDO) Atmospheric Imaging Assembly (AIA)  in EUV  on 2012 October 10
and  only a part of it, the pillar, which is a foot or barb of the prominence, with the Hinode Solar Optical Telescope (SOT) (in Ca II and  H$\alpha$ lines), Sac Peak (in H$\alpha$, H$\beta$ and Na-D lines),  THEMIS (``T\'elescope  H\'eliographique pour l' Etude du Magn\'etisme et des Instabilit\'es Solaires'') with the   MTR (MulTi-Raies) spectropolarimeter (in He D$_3$ line).
%  andThe small fields of view of SOT,  Sac Peak and the MTR  are  centered on a large, pillar-shaped prominence footpoint  extending towards the surface. This feature  appears in the full-disk AIA 304 \AA\  filtergram, as  a large quasi-vertical  pillar with material flowing horizontally on each side. 
 The THEMIS/MTR data indicates that the magnetic field in the pillar is essentially horizontal  and the observations in the optical   domain show a large number of horizontally aligned features on a much smaller scale than the pillar as a whole. The data is consistent with a model of  cool prominence plasma trapped in the dips of horizontal field lines.
The SOT and Sac Peak data over the 4 hour observing period  show  vertical oscillations  appearing as  wave pulses. These pulses, which include a Doppler signature, move vertically, perpendicular to the field direction, along thin quasi-vertical  columns in the much broader pillar. The pulses have a velocity of propagation of about 10 km~s$^{-1}$, a  period about 300 sec, and a wavelength around 2000 km.
 We interpret these waves in terms of fast magneto-sonic waves and discuss possible wave drivers.

\end{abstract}


%
%
%=============================================================================================================
\section{Introduction}
\label{Sect:1}
In this work we have an unusually complete set of observations with which to analyze oscillations in a large prominence footpoint or barb. These allow us to combine information concerning both three dimensional motions and magnetic field to analyze these as fast magnetosonic waves moving transverse the prominence magnetic fields and disturbing the cool material collected in the magnetic dips. 

Periodic motions in solar prominences have been routinely observed since the first observations of such oscillations in the 1950s \citep[e.g.,][]{ramsey1965}. These oscillations are classified as large- or small-amplitude oscillations depending on the velocity of the periodic motions, faster or slower than 20 $\mathrm{km~s^{-1}}$ respectively.
Large-amplitude oscillations are related to flare activity which causes  almost the entire filament to oscillate, shaken by the energetic event. Studies of small-amplitude oscillations have revealed a wide range of periods ranging from seconds to hours \citep{oliver2002,arregui2012}.
We will concentrate on the literature concerning the small amplitude  motions with periods from a few minutes to 10 minutes, called short-period  oscillations (see the review of \cite{mackay2010}), that is relevant to our present work.   

Prominences at the limb exhibit such  short period oscillations  \citep{oliver1999}. 
\citet{tsubaki1986} and \citet{tsubaki1987} reported short period oscillations in Dopplershift  observations of limb prominences. 
Two-dimensional observations of filaments on the disk have revealed oscillations transverse to the direction of  fine structures of the filaments with a period around 200 sec \citep{thompson1991,yi1991}. Recent high  spatial resolution observations obtained with the SST show the existence of traveling waves with periods of 3 to 9 minutes in fibril-like structures  of filaments. The propagation of the waves are in the same direction as the direction of the prominence fine structures.
as the mass flows. 
\cite{okamoto2007} found that threads in a solar prominence  observed with Hinode/SOT underwent vertically oscillating motions with a period around 250 sec. Some threads oscillate all along the filament length.  With  Hinode/SOT  data,  \cite{ning2009} analyzed the oscillations of a quiescent prominence and found a  large variety of  oscillations both vertical and  horizontal with periods of 3 to 6 minutes.
 
 The trigger of the small-amplitude, short-period oscillations has not been detected so far. 
% Motions with periods from few minutes to 10 minutes  called short-period oscillations and
 Many authors claimed that the excitation is related with the 3- and 5-minutes photospheric and chromospheric oscillations. 
Solar $p$-modes are generated by turbulent convection beneath the photosphere leaking part of their power to the atmospheric layers: photosphere, chromosphere, and corona (3- and 5-minutes oscillation).
%\cite{tsubaki1986,tsubaki1987} reported short period oscillations in spectrograph observations of limb prominences. The periods obtained were 160s, 210s, 240s, 400s, 640s, and 830s. Additionally, the authors found that the 

Oscillations in prominences are usually interpreted as an external agent exciting periodic motions on the cool plasma. MHD waves can be propagating or standing. Propagating waves consist of periodic disturbances that propagate through the prominence plasma. In contrast, standing waves are confined in a region of the prominence because they are the normal modes of the system. Several theoretical models of prominence oscillations have been proposed. \cite{joarder1992a,joarder1992b} considered the whole filament as a vertical plasma slab, ignoring the thread fine structure, with an horizontal straight magnetic field. The influence of the gravity was considered negligible and the slab and corona plasmas are uniform. In this model the waves are considered trapped along the horizontal direction, and are allowed to propagate in the vertical direction. \citet{oliver1992,oliver1993} included gravity in the slab model which results in the curved field lines of the Kippenhahn-Schl\"{u}ter equilibrium model. The authors found that the gravity introduces  only a small shift on the normal mode frequencies and thus is not relevant in this kind of configuration. Prominence fine structure consists of threads that could influence the global prominence oscillations. \citet{Joarder1997a} studied theoretically the oscillatory spectra of an isolated prominence thread. Later, \citet{diaz2006} included this fine-structure and studied fast MHD modes of a periodic, Cartesian multi-threaded model. The authors found that the modes behave as propagating modes of an homogeneous prominence with small-scale details due the fibrils.

In this paper we report the observations of oscillations in a quiescent prominence observed on the limb obtained by several instruments on the ground (THEMIS, Sac Peak)  and  in space (Hinode/SOT).  We describe the campaign and instrumentation in Section 2. In Section 3 we discuss the observations and present the results concerning the horizontal magnetic field measured in the prominence and the transverse oscillations. In Section 4 we discussed a possible model to explain the observed oscillations.

\section{Campaign and instruments}

The observations were taken during an international campaign organized around Hinode Observation Plan (HOP)
 219. Hinode/SOT observed in H$\alpha$  and in Ca II H lines  between 14:04 UT and 18:09 UT.  Over a similar time period (14:17-19:37 UT) the Dunn Solar Telescope (DST) at Sacramento Peak Observatory  (Sac Peak) was used to obtain spectra in the H$\alpha$, H$\beta$ and Na-D lines. THEMIS (``T\'elescope  H\'eliographique pour l' Etude du Magn\'etisme et des Instabilit\'es Solaires'' in Canary Islands) with the MTR (MulTi Raies) spectropolarimeter  observed the prominence  during all the day, obtaining four full data sets. The best data were collected  from 10:44-15:30 UT. SDO/AIA  with the filters of 304 \AA\, and 193\AA\ and STEREO-A EUVI at 195~\AA\  are used to supply context for the observations. 

\subsection{THEMIS}

The THEMIS/MTR instrument \citep{LARS00} was used to do spectropolarimetry of the He D$_3$ line in the observed prominences. The spectrograph slit was
oriented parallel to the local limb. This direction defined subsequently the sign of the linear polarization: positive Stokes Q means parallel to the slit and,
in consequence, parallel to the local limb. The double-beam polarimetry we performed required the use of a grid mask that presented us with three segments 15.5\arcsec\ wide along the slit, 
but masked regions of 17\arcsec\ between each slit. The masked regions allowed us to obtain a double image with opposite polarization, but it also meant that in order to get 
a continuous covering of the prominence along the slit we had to scan also in this direction by one step of 15\arcsec. This extraordinary scan along the slit is
the reason for the jumps and dark lines in the data from this instrument presented below. These artifacts are, however, greatly compensated for by the high
quality of the polarimetry produced by this observing mode. 

In addition to that scan along the slit a more usual scan perpendicular to the slit was 
made with steps of 2\arcsec\ from the limb to the top of the prominence. Altogether typical fields of view of $120\arcsec\times40\arcsec$ were covered in about one hour with 
single exposure times of 2 seconds per Stokes parameter and scan position. Full polarimetry with beam-exchange was done with a modulation cycle of 6 images, spanning the three Stokes parameters with either positive or negative
sign measured in every beam, and the simultaneous double beam measuring the opposite sign. Each Stokes parameter is thus measured in the same camera pixel at 
two different times and in two different pixels at the same time. This symmetry of measurements results in a reduction in the systematic errors  to a fourth order perturbation 
of the signal and high quality measurements. Each cycle was repeated five times to increase S/N ratios.

\subsection{Hinode/SOT H$\alpha$ and \ion{Ca}{2}}

The Hinode \citep{kosugi_07} SOT \citep{tsuneta_08,suematsu_08} consists of  a 50-cm diffraction-limited Gregorian telescope and a Focal Plane Package including the narrowband filtergraph (NFI), the broadband filtergraph (BFI), the Stokes Spectro-Polarimeter, and Correlation Tracker (CT). For this study, images were taken with a 30 sec cadence in both  the \ion{Ca}{2} H line at 3968.5~\AA\ using the BFI and at line center in the H$\alpha$ line at 6562.8~\AA\  using the NFI. The \ion{Ca}{2} images have a pixel size of 0.109\arcsec, with a field of view of 112$\times$ 112\arcsec, while the H$\alpha$ images have a pixel size of 0.16\arcsec and a field of view 164$\times$ 164\arcsec.

\subsection{Sac Peak Dunn Solar Telescope}\label{sec:dst-observations}
At the DST  at Sac Peak we used the Universal Birefringent Filter (UBF) to observe
filtergrams of H$\alpha$ at line center and line center $\pm 0.5$~\AA\ and $\pm 1.0$~\AA,
H$\beta$ at line center and line center $\pm 0.5$~\AA\ and Na-D at line center and
line center$-0.25$~\AA. The field of view is about $173\times$173\arcsec. In our analysis
we focused on the H$\alpha$ data. a full scan of the 5 H$\alpha$ line positions took 10 seconds and
time between images at H$\alpha$ line center was 22 to 25.6 sec. Pixel size is 0.17\arcsec, with resolution during our observations near 1\arcsec.

The full reconstructions of the H$\alpha$,  H$\beta$  profiles have not
been
done yet because it would require us to cross correlate the
non-simultaneous
images obtained in the 3-5 points in the profiles and then fit with a
Gaussian. This would allow to us to get quantitative estimates of the
Dopplershifts. It is out of the scope of the present paper. We use only
movies obtained by computing the difference of intensities of two
symmetric points in the profiles, a proxy for the Dopplershifts.

%Sac Peak  observations consisted in  filtergrams in H$\alpha$,  H$\beta$ and Na-D lines at line center  and in four points in the wings.   
%In each point of the field of view,  it is possible to compute the Doppler shift by reconstructing the line profile and measuring the displacement of the profile relatively to a mean profile obtained over the all prominence.  The calibration in wavelength is not accurate and only qualitative Doppler shifts can be derived.  \taknote{I think it is more that the absolute calibration is not known - is there a reason to think the relative wavelength scale is wrong or cannot be translated into km~s$^{-1}$?}

\section{Observations and Data Analysis}
Our observations of the prominence oscillations were done from 14:00 UT -18:00 UT on 2012 Oct 10.
The prominence was observed on the western limb of the Sun at a position angle of about 256$^\circ$.

\begin{figure}[!ht]
	\centering
\includegraphics[width=8cm]{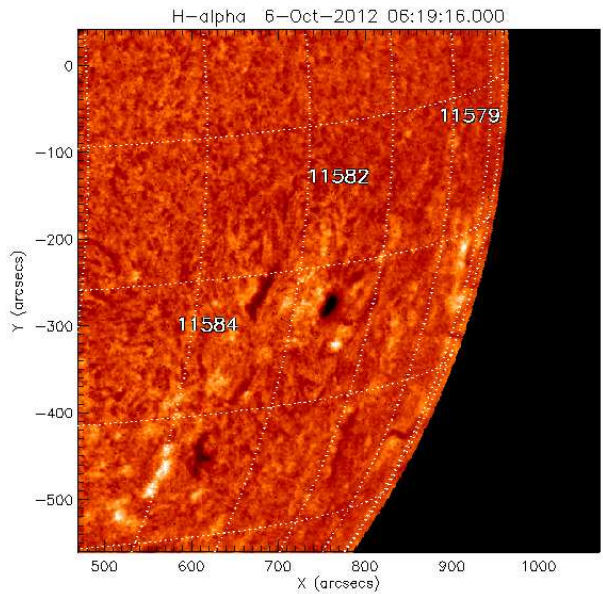}
\caption{H$\alpha$ Filament from BBSO on 2012 October 06   between the active regions NOAA 11582 and NOAA 11584.
\label{Fig:filament}}
\end{figure}

\begin{figure}[!ht]
	\centering
\includegraphics[width=7.8cm]{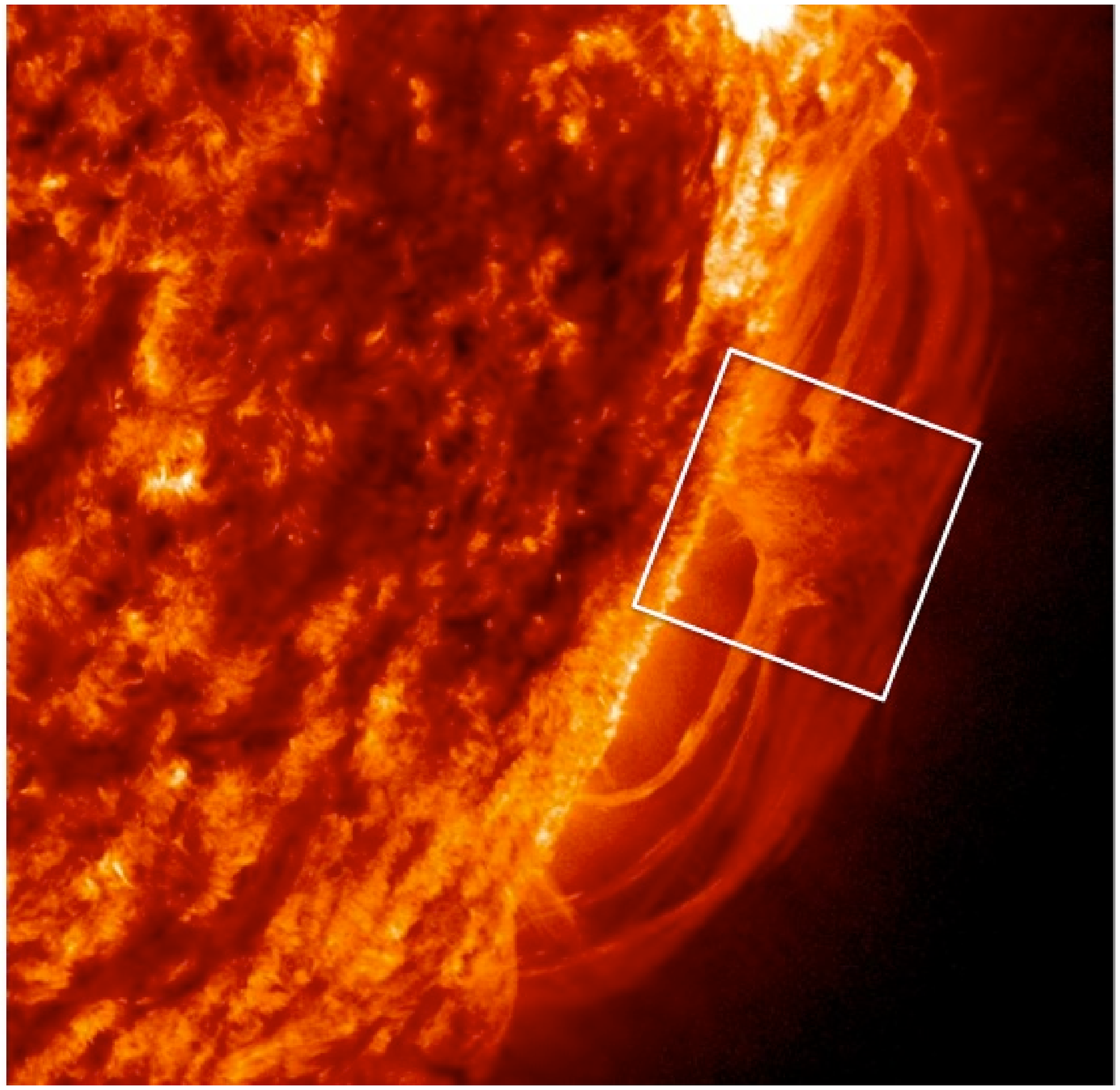}
\includegraphics[width=6cm]{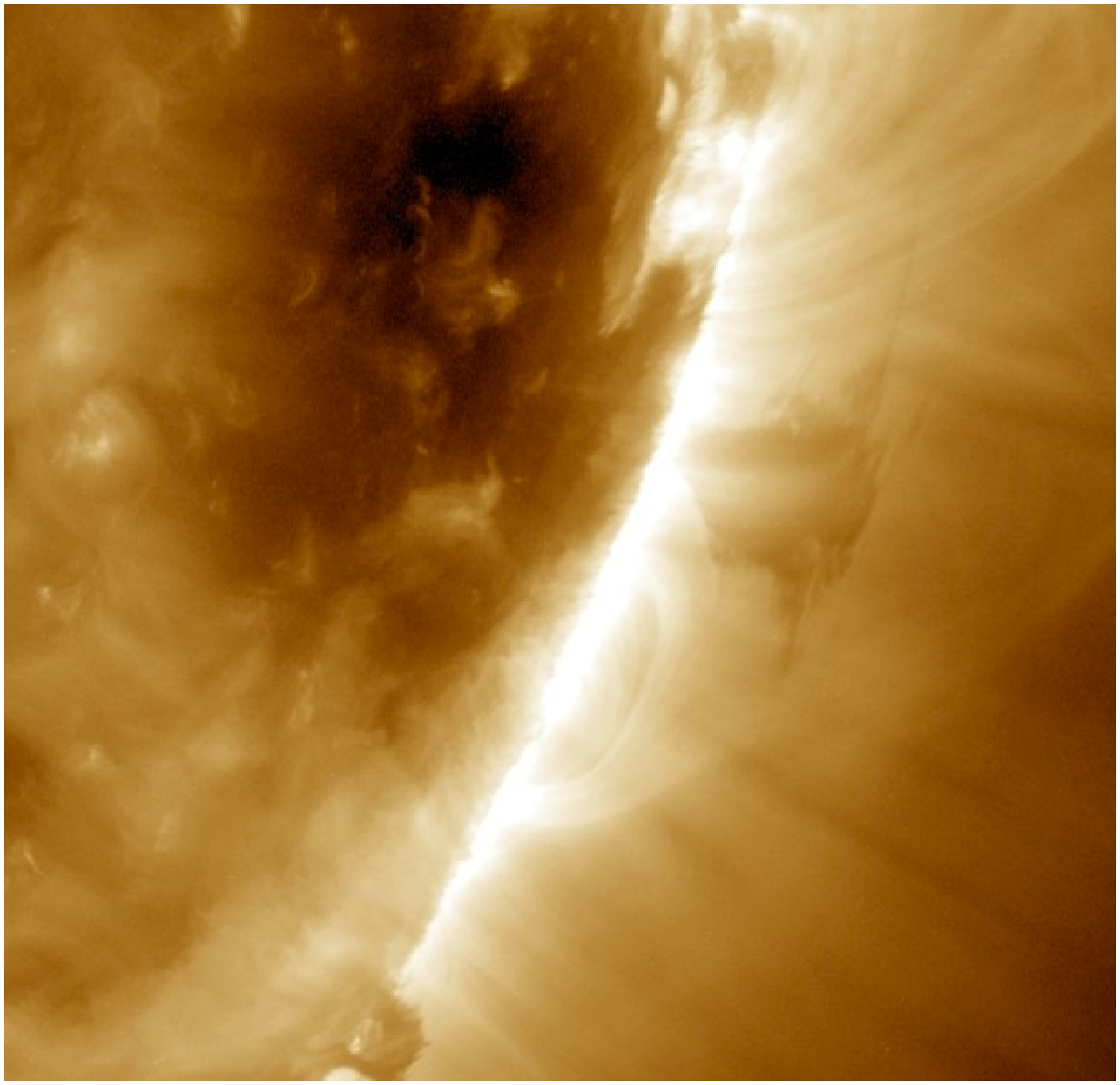}
\caption{(a)  AIA 304 \AA\, prominence observed on 2012 October 10. The box represents the field of view of Hinode and Sac Peak  (b)  
AIA 193 \AA\,  prominence on 2012 October 09.  The box is approximately the field of view of  Hinode/SOT. \label{Fig:AIA304}}
\end{figure}

\begin{figure}[!ht]
\centering\includegraphics[width=10cm]{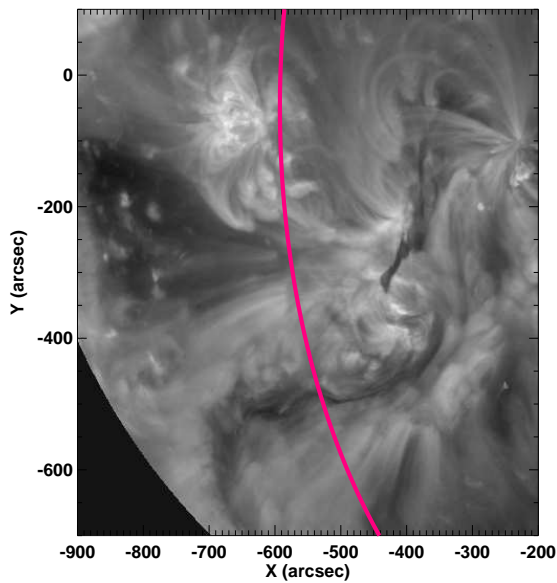}
\caption{(a) STEREO/SECCHI/EUVI-A image of the prominence observed in the 195~\AA\ band at  2012 Oct. 10 17:00:30~UT.
The pink curve shows the limb of the Sun as seen from Earth and thus also SDO and Hinode. At this time STEREO-A was at a separation of 126.18$^\circ$  ahead of Earth in its orbit.
\label{Fig:EUVIA195}}
\end{figure}

The prominence was observed  as a filament a few days before the main observations.
On October 6 only two portions of the H$\alpha$ filament are visible  in a channel between two 
active regions (Figure \ref{Fig:filament}). The filament is oriented North-South along a meridian. The prominence is well observed on  October 9 and 10  in SDO/AIA filters. As seen in AIA 304~\AA\  it consists of a central pillar with  arcades on both sides from which material is flowing horizontally, mostly outwards away from the central pillar in the plane of the sky (Figure \ref{Fig:AIA304}). On October 9  the central pillar is already observed  in absorption as a dark region over the limb in AIA/193~\AA. On October 10  at the time of Hinode observations bright loops are in front of it and mask  the dark area in 193~\AA.  By this time the prominence was already substantially behind the limb as is shown in the image from STEREO-A's EUVI imager in Figure \ref{Fig:EUVIA195}.

\begin{figure*}[!ht]
\includegraphics[width=10cm]{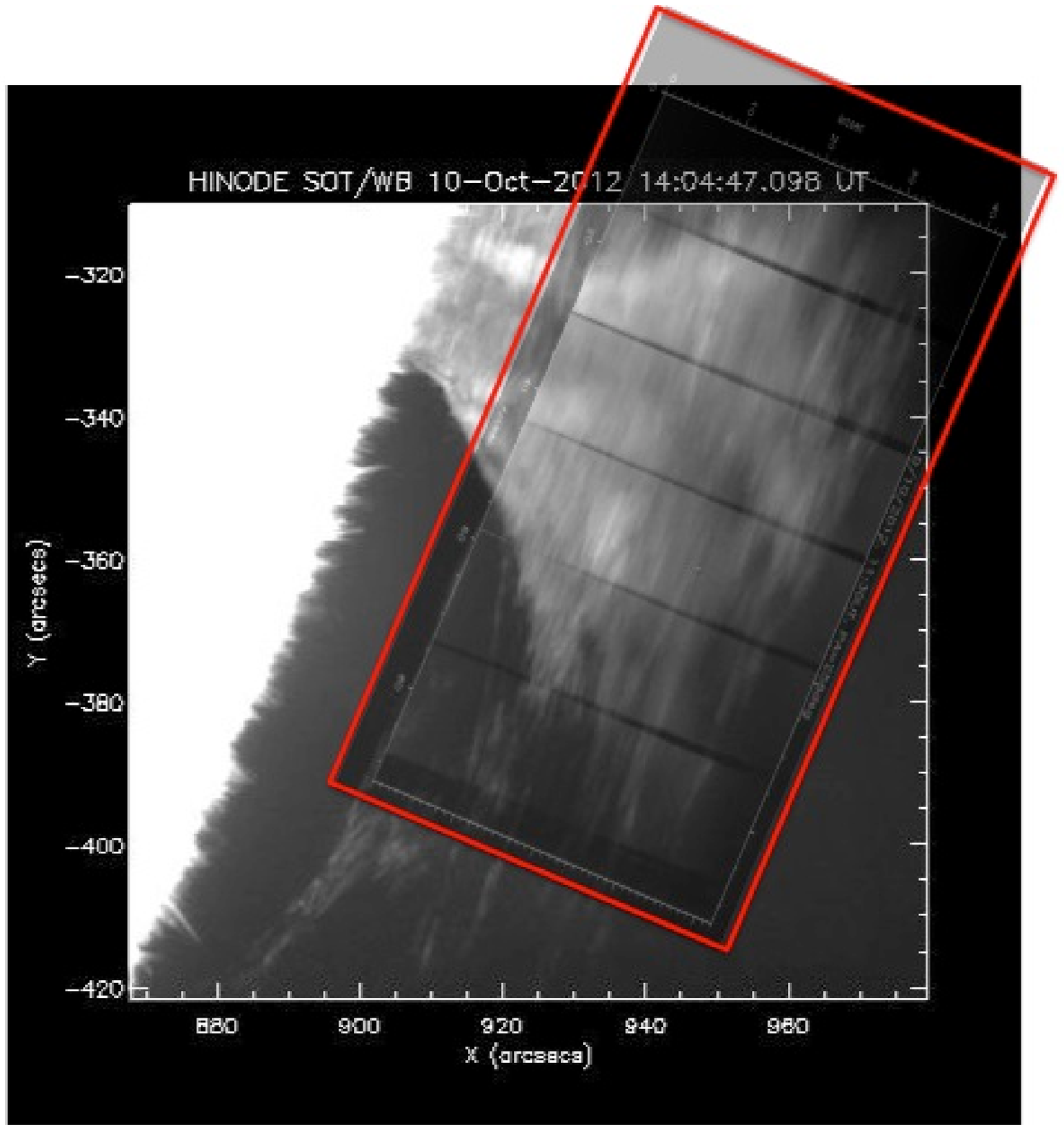}\hspace{-2cm}\includegraphics[width=10cm]{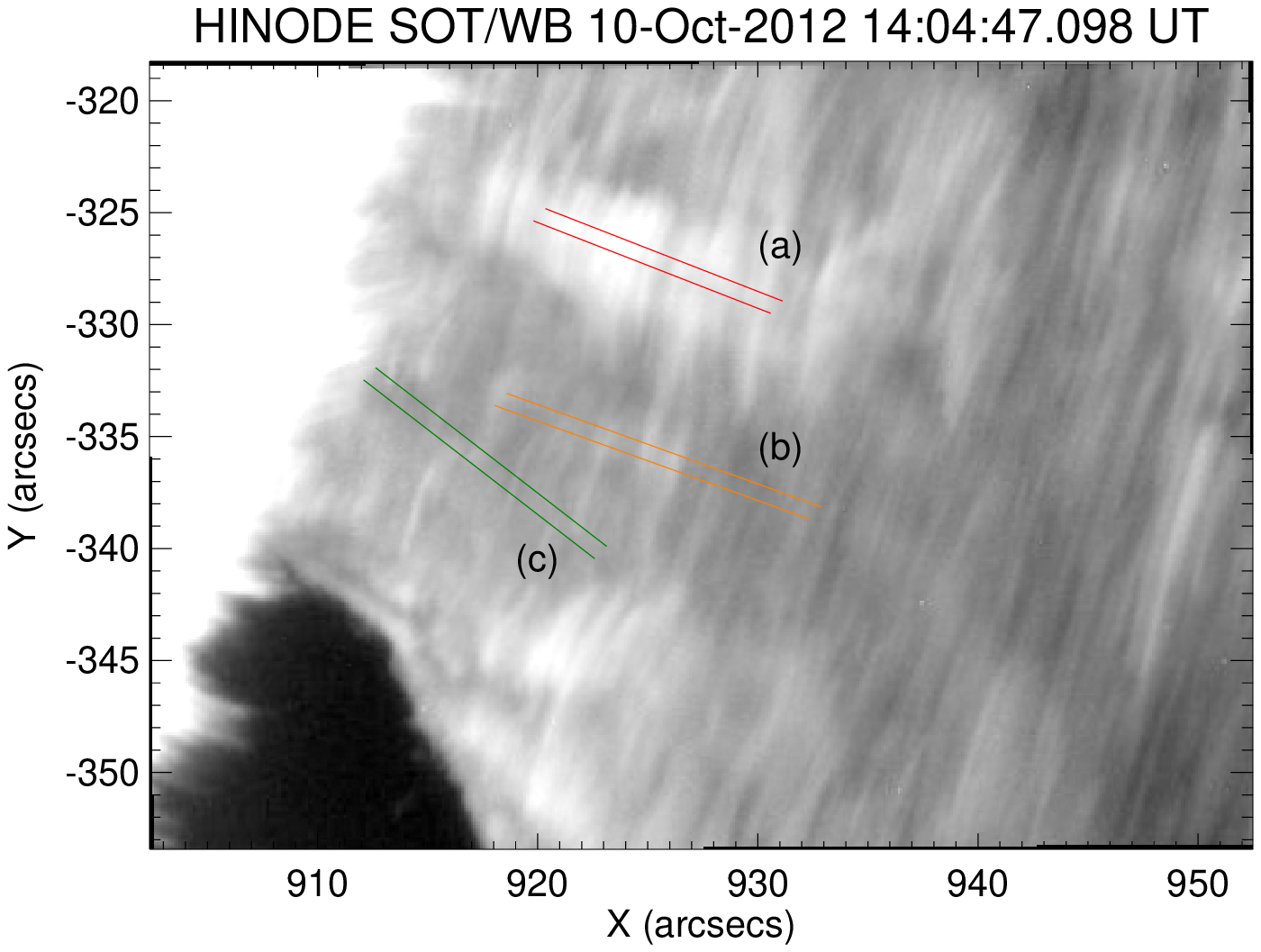}
\caption{{\it left panel:} Hinode prominence in  the \ion{Ca}{2} H line at 14:00 UT on October 10 2012. 
The box is approximately the field of view of THEMIS/MTR.
{\it right panel:} Hinode/SOT image showing the location of the three 5 pixel, 0.55\arcsec\ wide
slits along which the intensity of the oscillation shown in Fig.~\ref{Fig:sot2} was calculated. 
The labels (a), (b), and (c) correspond to panels in Figs.~\ref{Fig:sot2} and \ref{Fig:sot3}.
\label{Fig:sot1}}
\end{figure*}

Hinode and the ground based instruments (Sac Peak and THEMIS) have smaller fields of view and observed mainly the central pillar (Figure \ref{Fig:sot1}).   

%   Figure 1 (a)  AIA/304 prominence observed on October 10, 2012. The box represents the field of view of Hinode and Sac Peak  (b)  Hinode  prominence in 
%Halpha (or Ca II)  at  UT ; The box is the field of view of THEMIS/MTR
 
\subsection{Wave Observations}
\label{Sect:WaveObs}
% \subsection{Analysis and Data Reduction of SOT data}
Hinode/SOT observed the prominence  from 14:04-18:09 UT.
The field of view of SOT is centered on the central foot of the
prominence. We see mainly a large, broad pillar with bright, relatively
narrow ($< 5\arcsec$) columns consisting of horizontal features and
lateral extension at the top.
For our measurements here we analyzed the \ion {Ca}{2} data, although
the waves were apparent in the H$\alpha$ data as well. The
Hinode \ion{Ca}{2} data were reduced using the standard SOT reduction
software (fg-prep) in SolarSoft.

The SOT/CT  \citep{shimizu_08} allows fixed tracking of regions on the
disk, but for regions on the limb there is a slow drift of the target
through the field of view. We correct for this drift using manual
corrections for large jumps associated to re-pointings and a cross
correlation targeted at the limb and the side of the large scale
prominence foot-point. The resulting image series is sufficiently stable
that remaining jitter and drift do not affect the motions and changes
measured for this paper.

\begin{figure}
\centering\hspace{-2cm}\includegraphics[width=10cm]{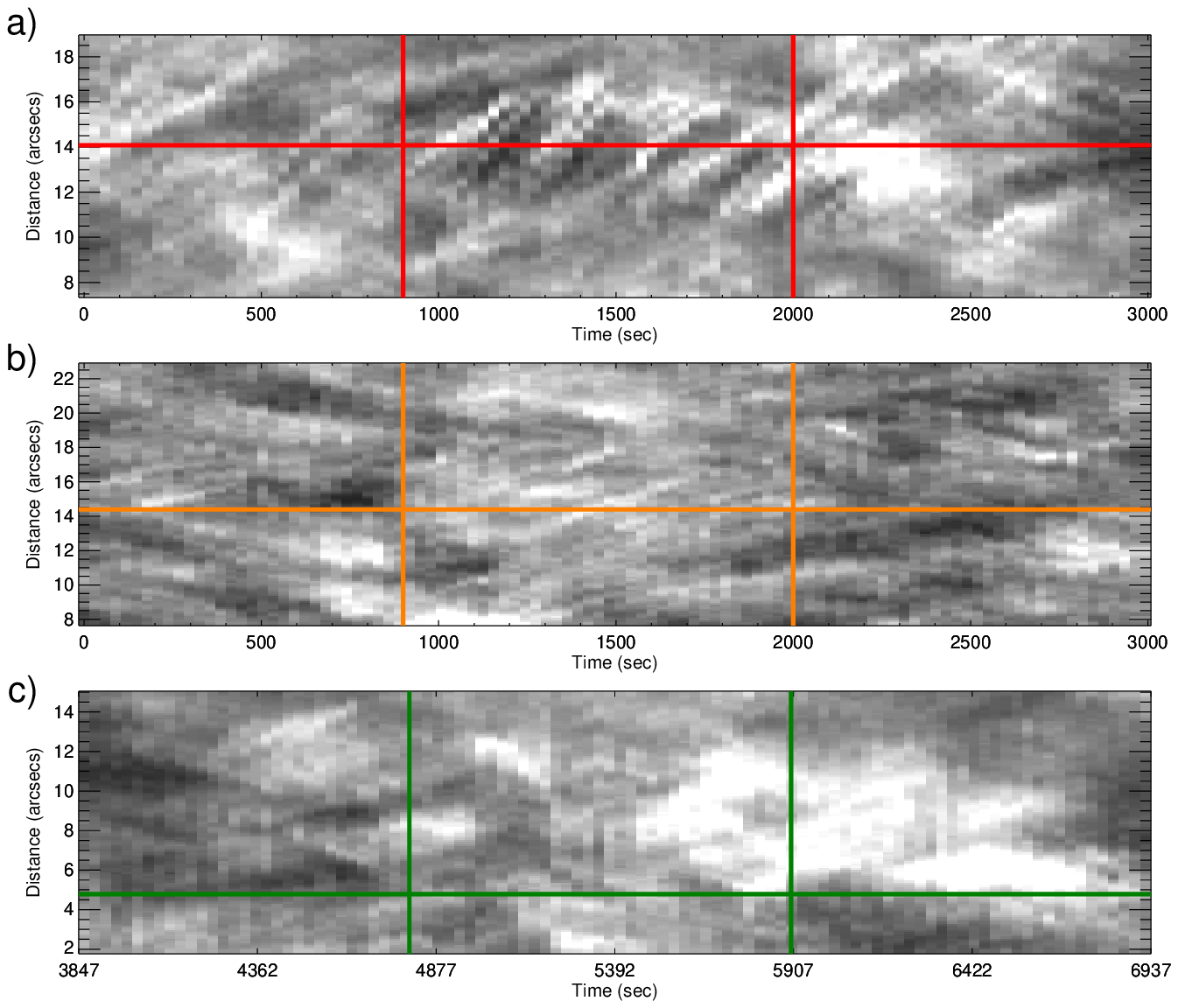}\\
\centering\hspace{-2cm}\includegraphics[width=10cm]{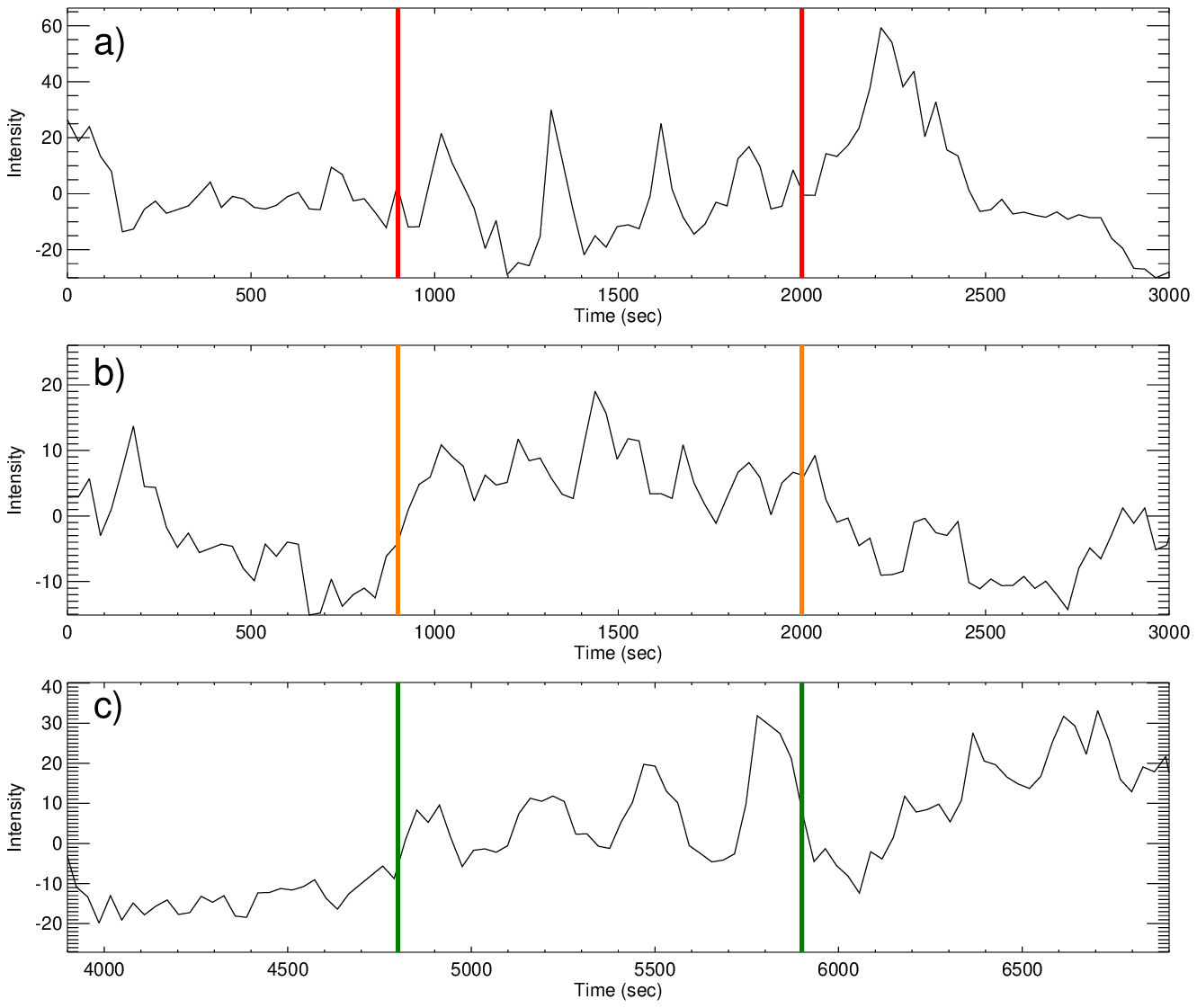}
\caption{{\it top panels} (a) (b) (c) Intensity maps as a function of time along the three slits shown in Figure
\ref{Fig:sot1}. Cuts of the intensity was made along the horizontal line, shown in {\it bottom panels} (a) (b) (c ). The
vertical lines represent the range of times of the data used for the periodogram
results in Fig.~\ref{Fig:sot3}. Long term trends were subtracted from the intensities resulting in negative values.
\label{Fig:sot2}}
\end{figure}

\begin{figure}
\centering\includegraphics[width=9cm]{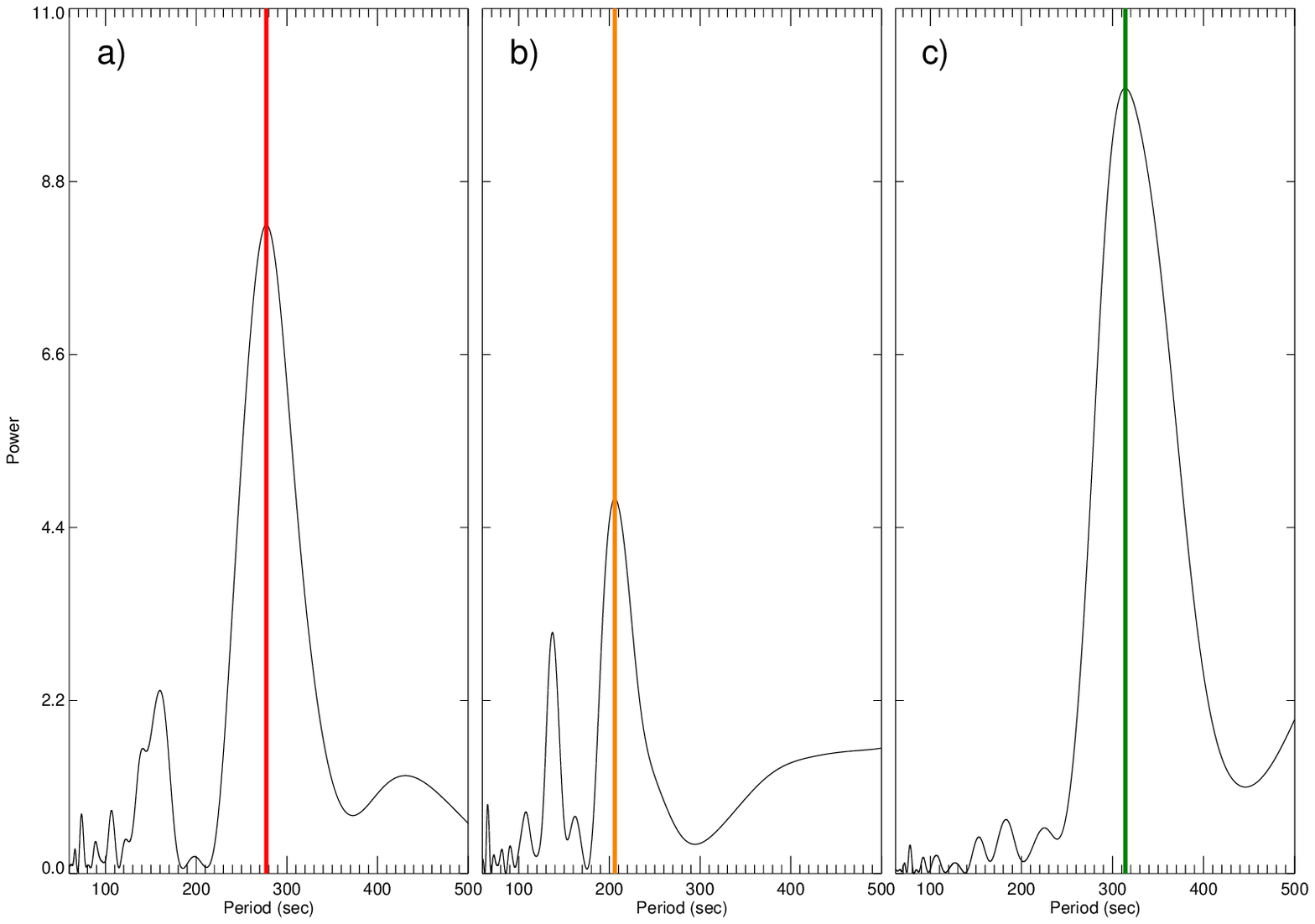}\\
\centering\includegraphics[width=9cm]{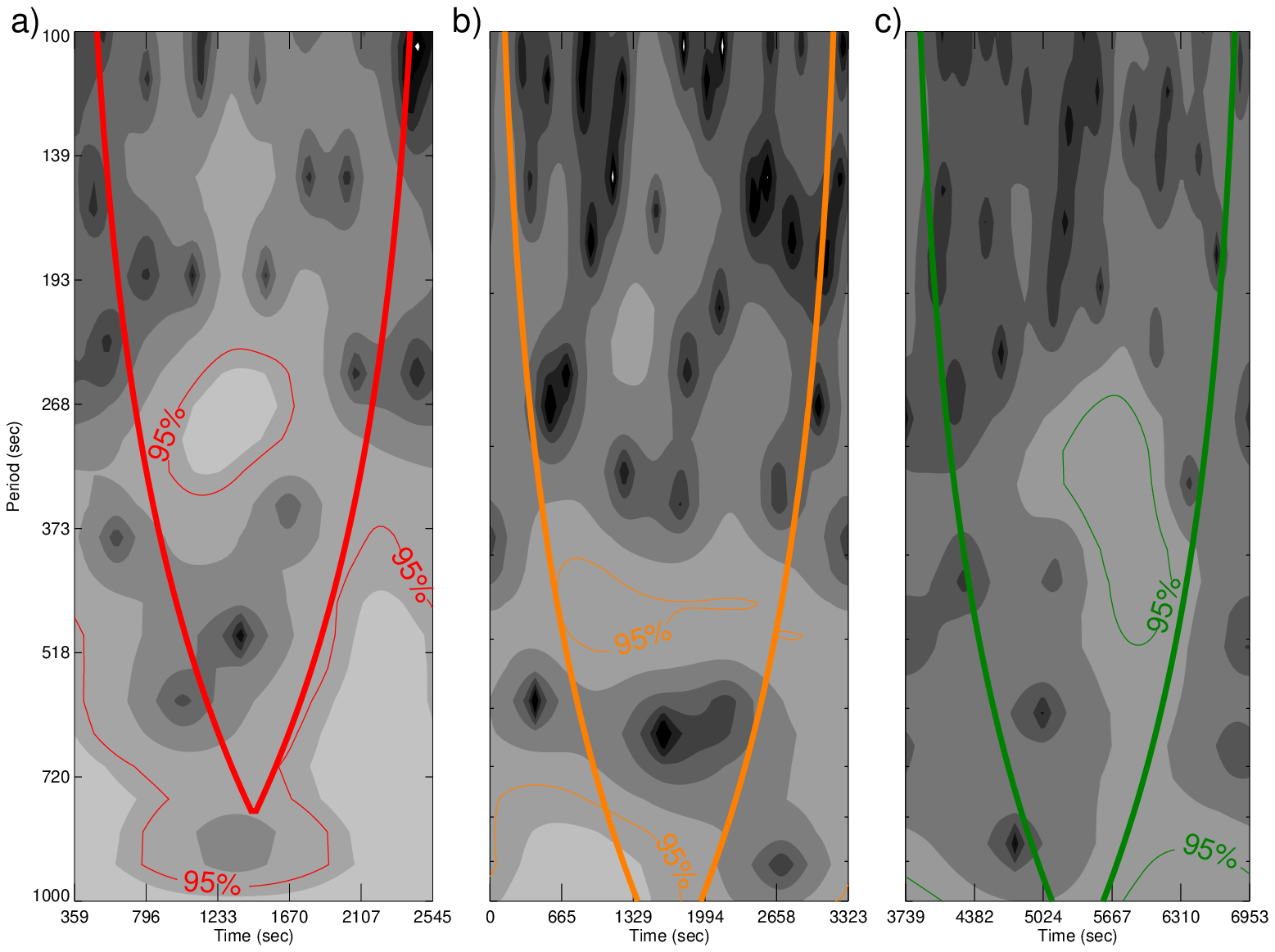}
\caption{ {\it top panels} (a) (b) (c) Periodograms of the intensity cut shown in fig. \ref{Fig:sot2}
The peak power
is located  respectively at 277 sec ($\pm$ 50 sec),  205 sec ($\pm$54 sec) and 314 sec ($\pm$125 sec). 
{\it bottom panels} (a) (b) (c) Confidence regime for the
periodograms, indicating that the period is staying approximately constant for the first two waves but many be changing for the third wave.
\label{Fig:sot3}}
\end{figure}

   The Hinode SOT images show many features which appear to be wave
pulses traveling roughly perpendicular to the solar limb in the narrow
columns. The analysis of the most clear of these motions is shown in
Figures \ref{Fig:sot1}, \ref{Fig:sot2}, and \ref{Fig:sot3}.
We integrated the intensity across a 0.55\arcsec\ (5-pixel) wide area
positioned across the oscillating region as shown in Figure
\ref{Fig:sot1} at three different locations, with slits labeled (a), (b), and (c).
A 50 minute long intensity slice is shown in Figure
\ref{Fig:sot2} (top panel a), corresponding to the red slit labeled (a).
This wave occurs between 1000-2000 seconds,
where all times are measured with respect to 14:04:47 UT. In
order to remove long term variations, the intensity data were fitted
with a quadratic function which was then subtracted from the intensities.
The distance between each intensity peak, measured at a given time, is a
measure of the wavelength of the oscillation, and is approximately 2000
km. The slope of the intensity peaks in Fig.~\ref{Fig:sot2} (top panel
a), corresponding to the upwards velocity of the moving features, is
approximately $10 \pm 4$ km~s$^{-1}$. The velocity corresponds to the phase speed of
the wave. Since the phase speed remains approximately constant, the wave
can be considered to be non-dispersive. In Figure \ref{Fig:sot2} (bottom
panel a). we plot the intensity as a function of time at an altitude of
14\arcsec\ above the photosphere. Four peaks and troughs are seen in
this time range. A Fourier analysis of this section of the intensity
cut, shown in Figure ~\ref{Fig:sot3} (top panel a), gives a wave period of
277$\pm$50 sec.  The wavelet analysis, shown in Figure \ref{Fig:sot3}
(bottom panel a),  indicates that the period of the wave remains
approximately constant for the duration of the wave.

Identical analyses were performed at two other different locations, as
shown in Figs. \ref{Fig:sot1}, \ref{Fig:sot2} and \ref{Fig:sot3} (panels
b and c). The wave measured by the intensity slice shown in orange and
panel b in Figs.~\ref{Fig:sot2} and \ref{Fig:sot3}, shows a similar
series of bright moving features between 900 and 2000 sec.  The speed of
the wave is measured to be approximately $5 \pm 3$ km~s$^{-1}$ radially
outward, and the wavelength is estimated to be
roughly 900~km. The intensity cut at an altitude of 5\arcsec\
above the solar limb is shown in Figure  \ref{Fig:sot2} (bottom
panel b).  The period of this wave is seen from Figure \ref{Fig:sot3}
(top panel b) to be consistently near $205 \pm 54$ sec.

For the case c (Fig.~\ref{Fig:sot1} green slit) from 4800-5900 sec
the wave is propagating downwards rather than upwards with a velocity of
-5$\pm 2$ km~s$^{-1}$ and a period of $314\pm 125$ sec  (Figs. \ref{Fig:sot2} and
\ref{Fig:sot3}  panels c) and may be decreasing over time. The
uncertainty has been determined by taking a vertical cut through the
confidence region of Figures  \ref{Fig:sot3} (bottom panels a, b and c  
respectively).

%\subsection{Dopplershifts Sac Peak}
    %Toot and Kucera
   The Dopplershift movies  in H$\alpha$ and H$\beta$ lines obtained from the Sac Peak data  show the existence of the same traveling waves at the same locations as the Hinode  \ion{Ca}{2} intensity observations. These  waves can be  detected both in intensity and in Dopplershift. 
   
\begin{figure}
\hspace{-1.5cm}\includegraphics[width=12cm]{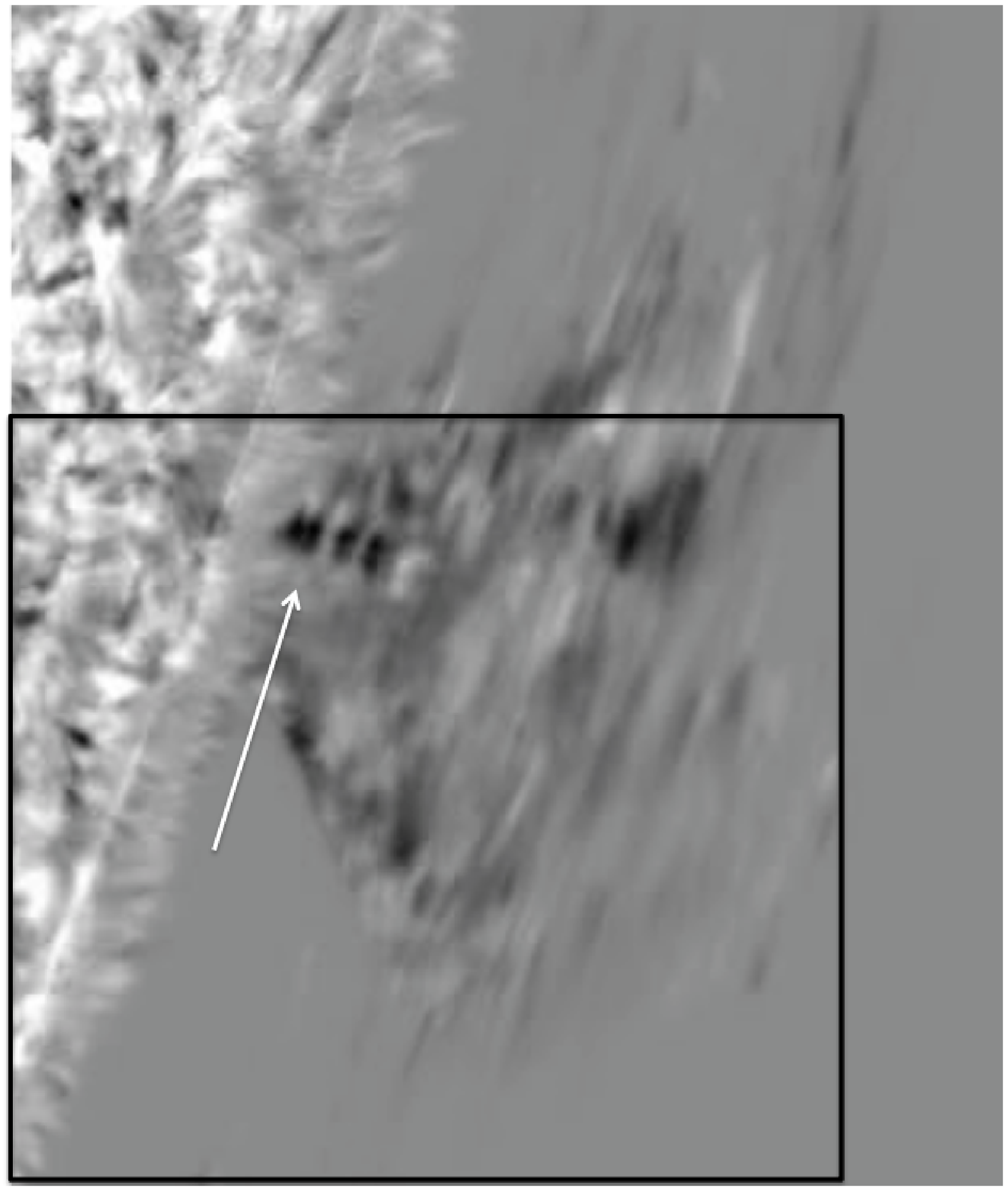}	
\caption{Dopplershifts  in the prominence derived from Sac Peak H$\beta$ observations  (173 x173\arcsec).  Note the large Dopplershift in the column  (similar to the column indicated by a red slit in the Fig.\ref{Fig:sot1} right panel) pointed by an arrow, the box is approximately the field of view  of Hinode/SOT.
%The dopplershift in the dark column in the prominence is estimated to be around -20 km~s$^{-1}$.
%\taknote{Would be good here to have an over plot of the Hinode field of view and maybe x and y axes showing arcsec}
\label{Fig:sac}}
\end{figure}

Figure \ref{Fig:sac}  shows a snapshot of the Sac Peak  observations of the prominence (H$\beta$ Dopplershift )  on Oct.\ 10 2012. The maxima of the Dopplershifts moving up in the column of the wave are redshifted compared to the whole prominence.  
%{\bf I suppress the value)}
%The relative velocity is estimated  to be $-20\pm10$ km~s$^{-1}$ compared to the whole prominence.
%, (a) intensity,b) Dopplershifts. The line is radial.
    
%  Variation of the Dopplershifts along the line, time slice evolution of Dopplershifts.

\subsection{Magnetic field vector}

 The raw data of the THEMIS/MTR mode was reduced with the DeepStokes  procedure \citep{LAARMSDG09}. Data reduction included flat-fielding, dark current and bias subtraction, wavelength 
calibration and, particularly, a careful handling of the polarization signals. The result of the data reduction are cubes of spectra of the He D$_3$ in intensity,
linear polarization (both $Q$ and $U$) and circular polarization, for all points along the slit and all positions of the double scan. S/N ratios are better than
$10^3$ at the core of the He D$_3$ in the central parts of the prominence for all three Stokes parameters. With these S/N ratios we get clear signals
of linear polarization as expected (these are a function of height above the limb, but linear polarization is expected at the level of $10^{-2}$ times the intensity and therefore 10 times
the noise level) but circular polarization is  seldom seen above the noise. Whether this circular polarization is due to Zeeman effect or the alignment-to-orientation 
transfer mechanism \citep{LAC02} those low signals already point to weak magnetic fields, a conclusion that will be confirmed by the inversion codes.

The Stokes profiles are fed to an inversion code based on Principal Component Analysis \citep{LAC02,Casini03} that efficiently compares the observed profile against
those in a database generated with known models of the polarization profiles of the He D$_3$. The  comparison is made pixel by pixel independently. The database used contains 90000 profiles computed as the emission of 
a single He atom in its triplet state modeled with the 5 levels of lower energy of the He triplet system. The atom is polarized by the anisotropic radiation of
the photosphere below the prominence at different heights (one of the free parameters of the model). Collisions are not  taken into account. The atomic
polarization of the He atom is modified by a single vector magnetic field with free strength, inclination and azimuth. The Hamiltonian of the atom includes all 
terms with its Zeeman sublevels splitting linearly with the magnetic field. We solve the density matrix  of the atom in statistical equilibrium; the solution 
contains all populations and quantum coherence, including 
atomic alignment and orientation, for all the levels involved in the He triplet atom model. The Hanle effect of every level is thus computed as well as the Zeeman 
effect. From the resulting populations and coherence we compute the polarization 
dependent emission terms,  in whatever direction we are observing. The scattering angle is thus a free parameter of the model too. Several million profiles thus computed are used to fill 
the database, keeping just those which are different enough among them and rejecting others so that the database fills as homogeneously as possible the space of 
possible profiles while keeping at a small size. 

After comparison of any observed profile with those in the database, the most similar is kept as the solution and the parameters of the model used in its computation 
are kept as the inferred vector magnetic field, height above the photosphere and scattering angle. Error bars are determined for those parameters as well by doing
some statistics on all other models which are sufficiently similar to the observed ones, though not as similar as the one selected as solution. It is important to
stress that although there is always one case in the database that is the most similar one to the observed one, this does not mean that it is a good fit to all
of the observed profiles. It is thus important to keep a measure of how similar they are and also to check that all conclusions on the magnetic field strength or 
orientation are based upon sets of profiles that really  correctly represent the observation.

\begin{figure}
	\centering
\includegraphics[width=7cm]{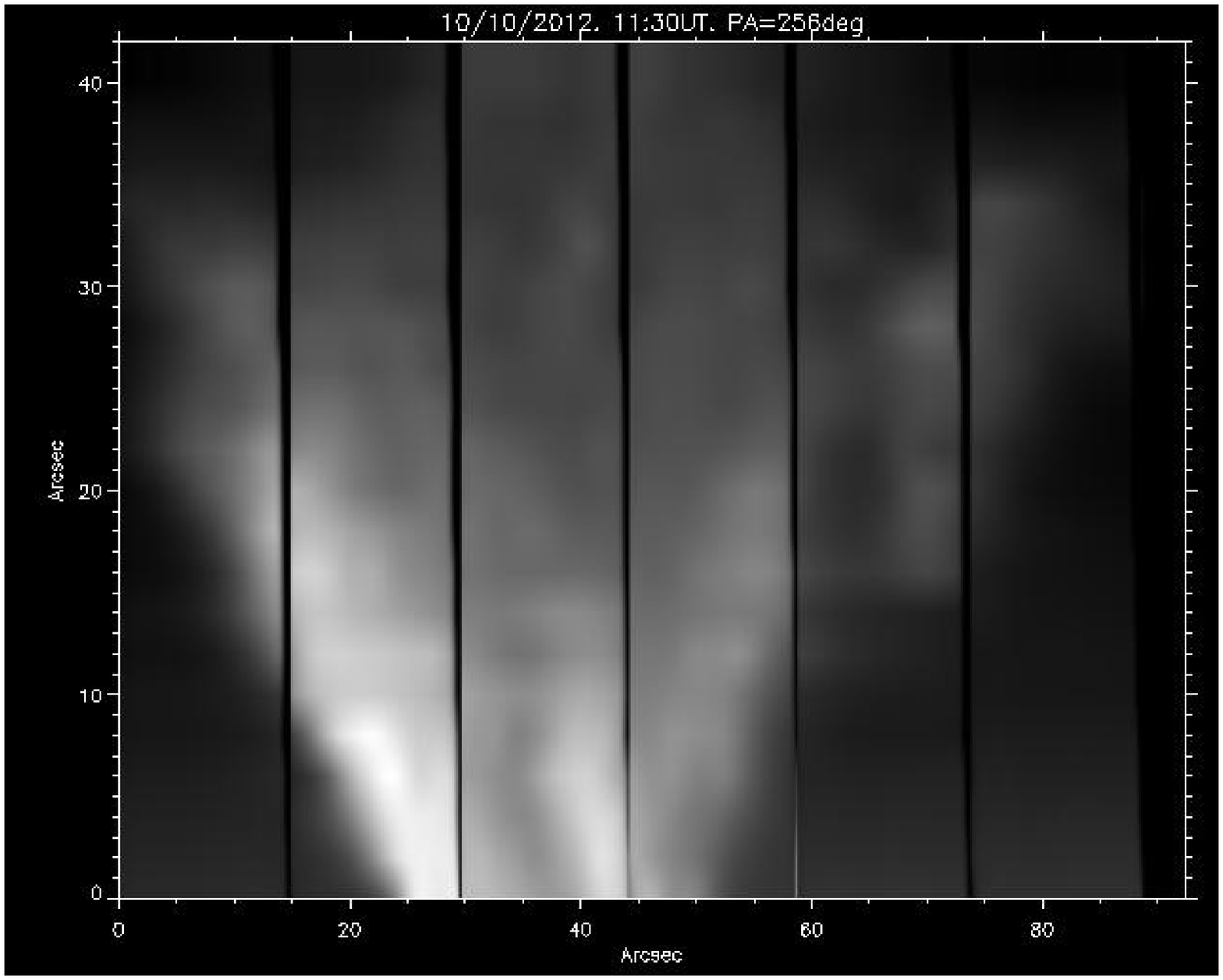}
\includegraphics[width=7cm]{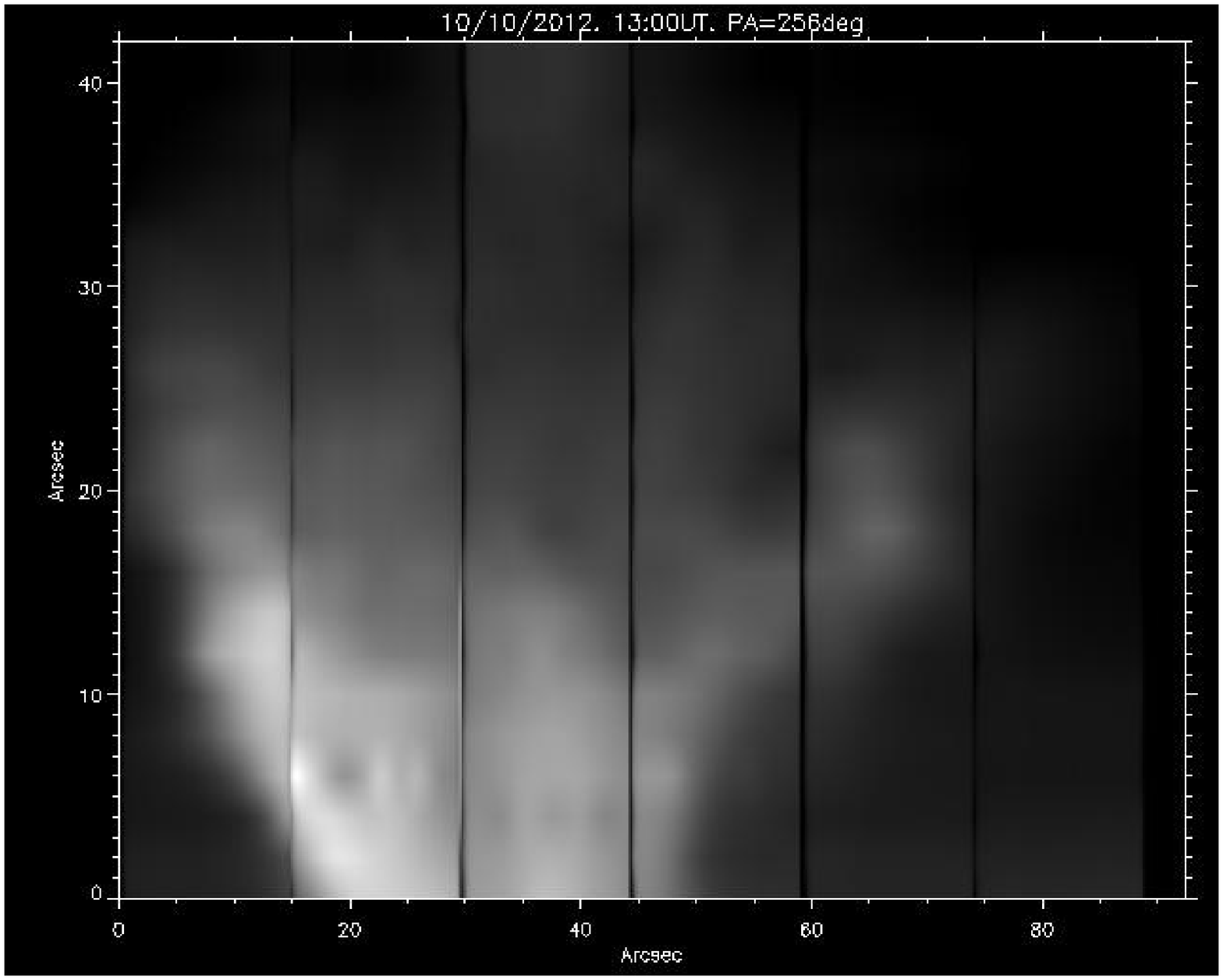}
\includegraphics[width=7cm]{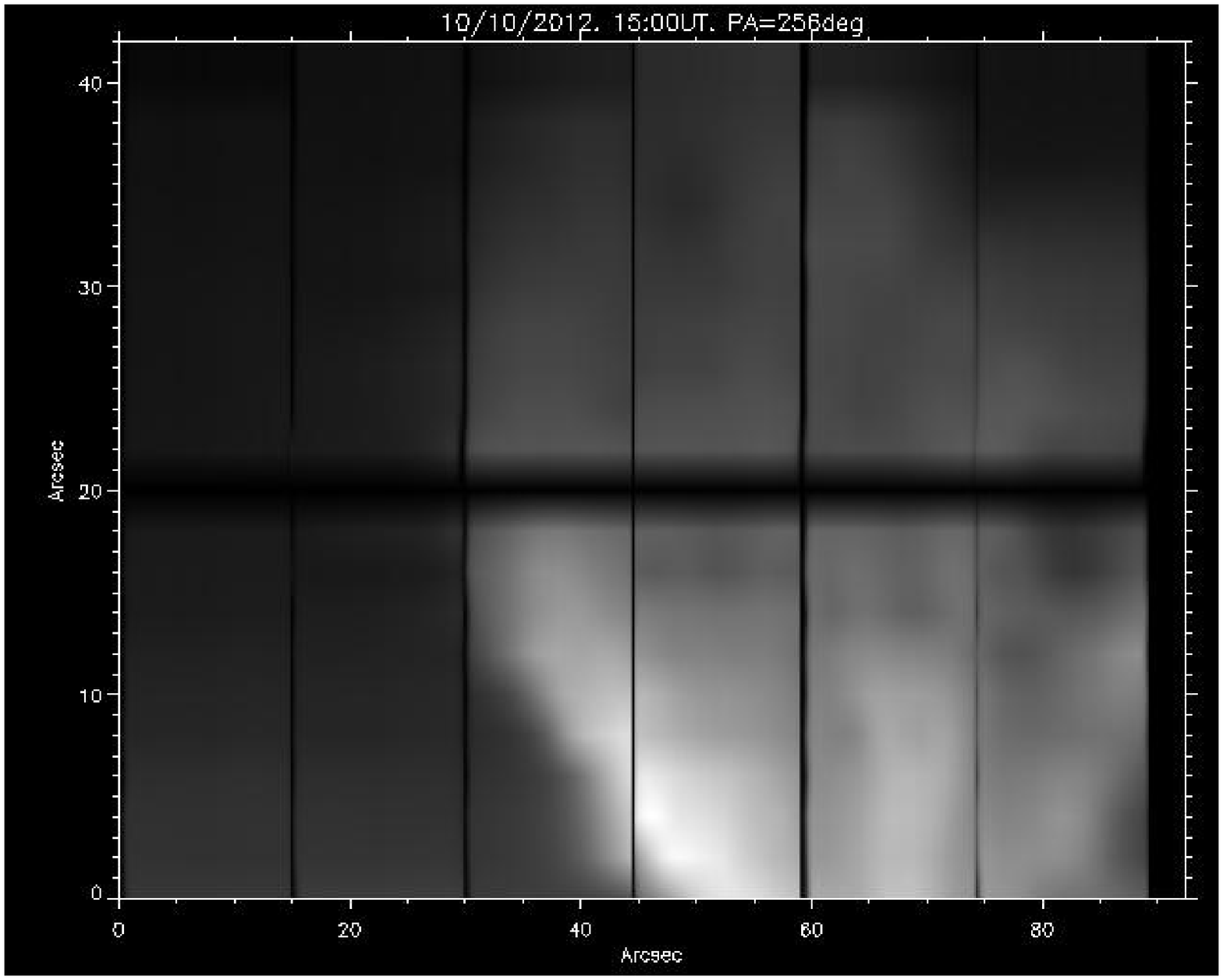}
\caption{THEMIS/MTR observations of the prominence in He D$_3$ line intensity: between (a) 10:44 and 11:52~UT, (b)  12:09 and 13:12~UT, and (c) 14:26 and 15:30~UT. The fields of view are $42\arcsec\times90\arcsec$. The fields of view are not the same. They have been shifted by a  few pixels towards the top of the prominence. The images are rotated so that the  limb is horizontal. The dark vertical lines are due to the grid mode of the observations. 
\label{Fig:themis1}}
\end{figure}

Figure \ref{Fig:themis1}  presents three  of the four  maps of the prominence intensity obtained on October 10   with  THEMIS/MTR in the He D$_3$ line.  

\begin{figure}
	\centering
\includegraphics[width=8cm]{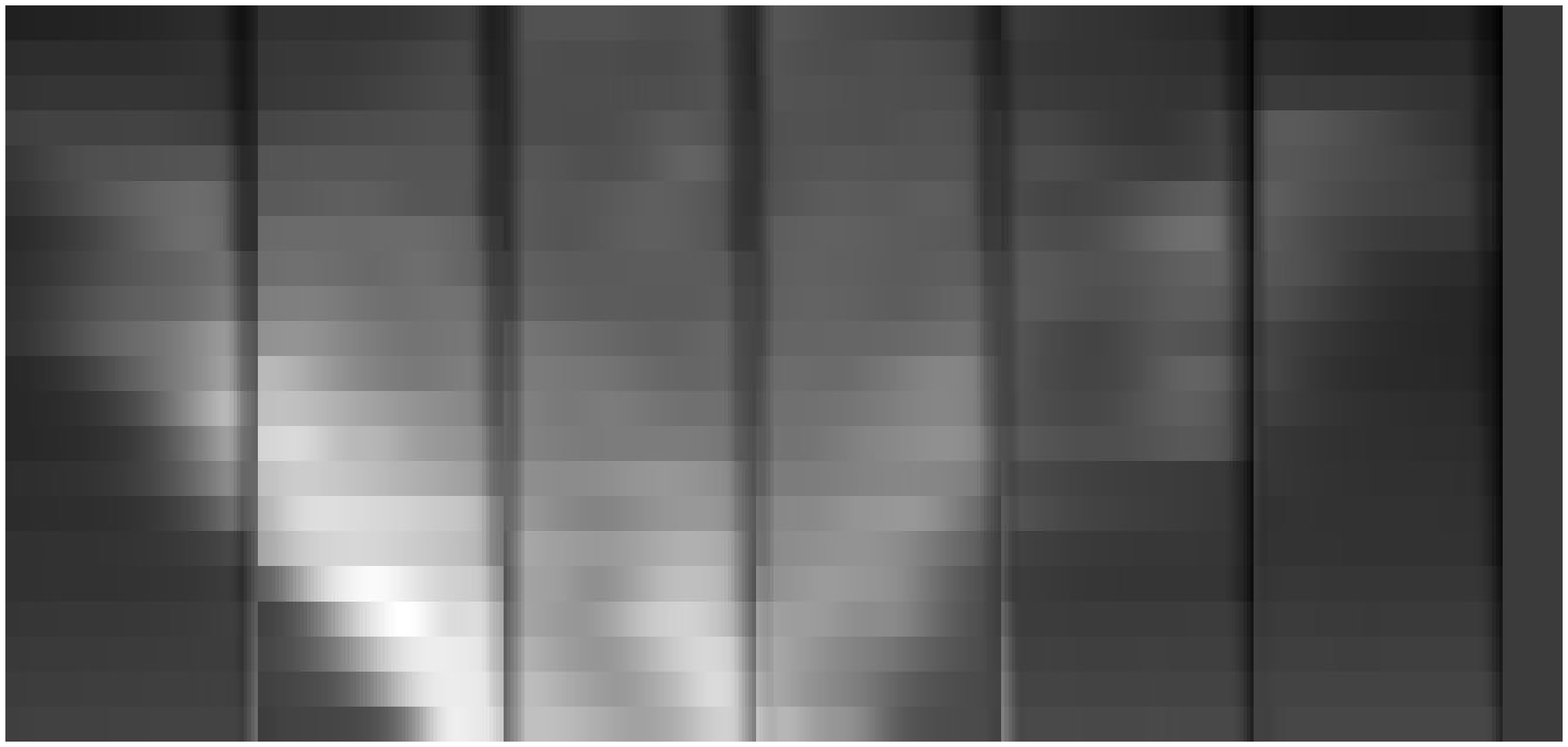}
\includegraphics[width=8cm]{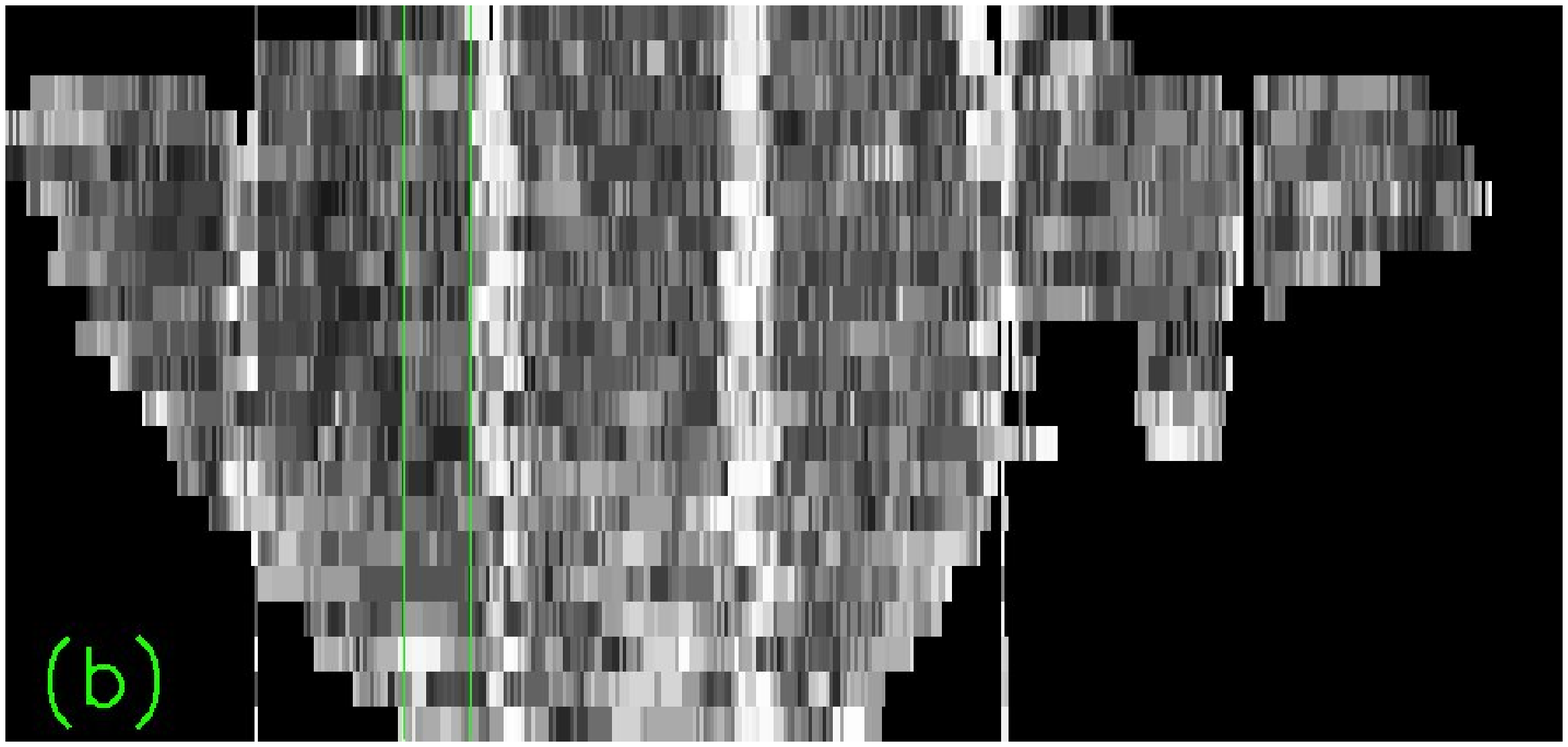}
\includegraphics[width=8cm]{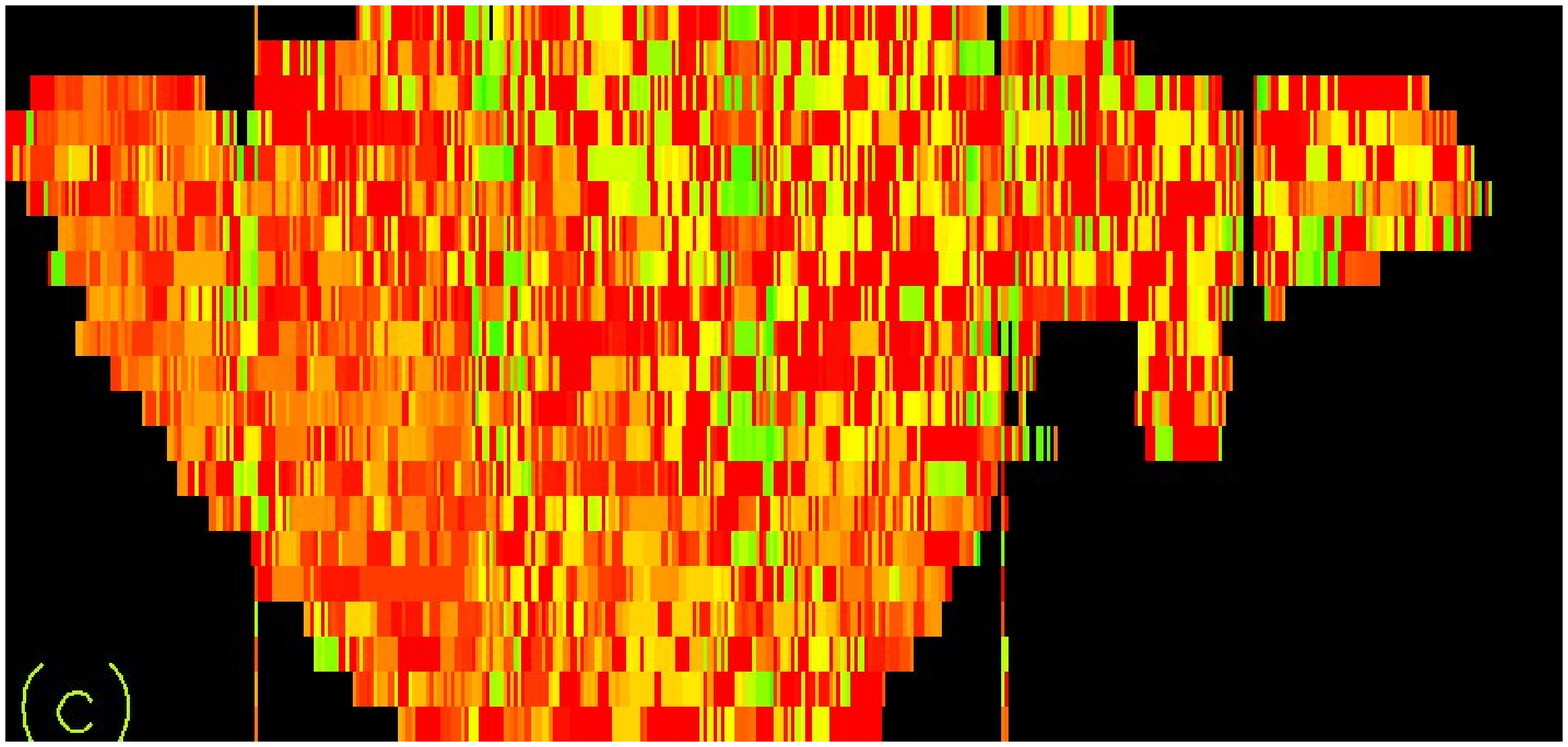}
\includegraphics[width=8cm]{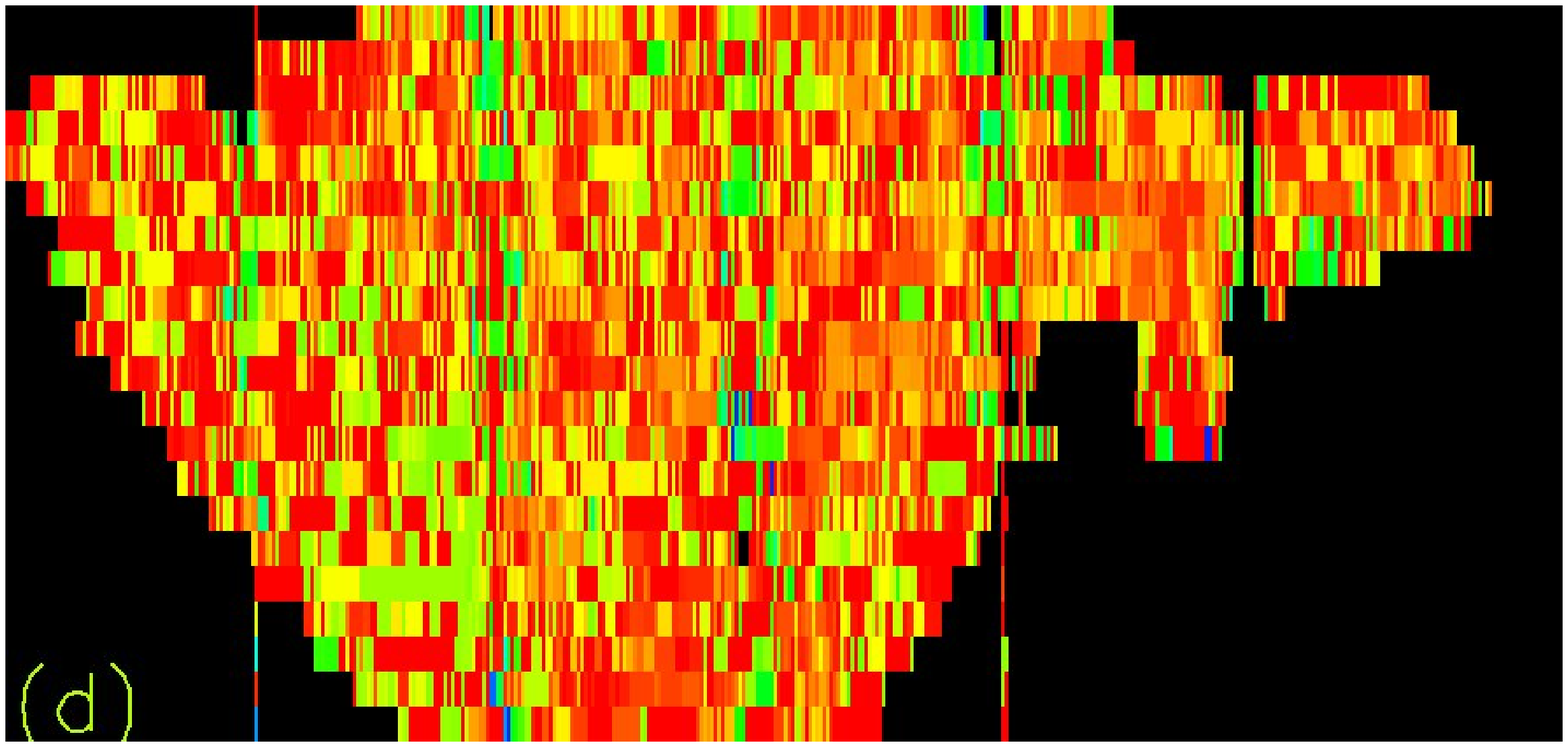}
\includegraphics[width=5cm]{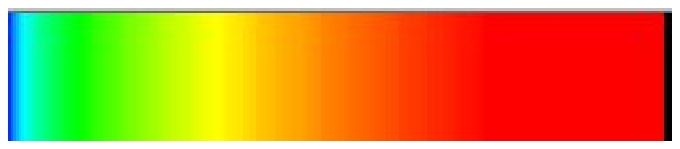}
\caption{THEMIS/MTR observations of the prominence in He D$_3$ line : (a) Intensity  between 10:44 and 11:52 UT  (b) Magnetic field strength (c) Inclination, (d) Azimuth.
% ; The arrow (vertical line)  shows the region of  the cuts in the next Figure.  
The color chart refers from  0$^\circ$ to +180$^\circ$ (left to right) .  Orange means around 90$^\circ$. The inclination is measured from  the vertical.  All the orange  pixels   in the inclination map  (on the left ) show that the field direction is mainly horizontal. The azimuth is mainly  around 110$^\circ$, so that the field direction is  directed approximately (within about 30$^\circ$) parallel to the plane of the sky 
\label{Fig:themis2}}
\end{figure}

Figure \ref{Fig:themis2}  presents the maps obtained after inversion  of the Stokes parameters recorded in the He D$_3$  line with THEMIS/MTR  : (a) Intensity, (b) Magnetic field strength (c) Inclination, (d) Azimuth.  The angle origin  of inclination is the local vertical, the  origin of the azimuth is  a plan containing the line of sight and the local vertical. We see that the brightest parts of the prominence have an inclination of 90$^\circ$ which means that the magnetic field  in these bright columns is horizontal. 

\begin{figure*}
	\centering
\includegraphics[width=16cm]{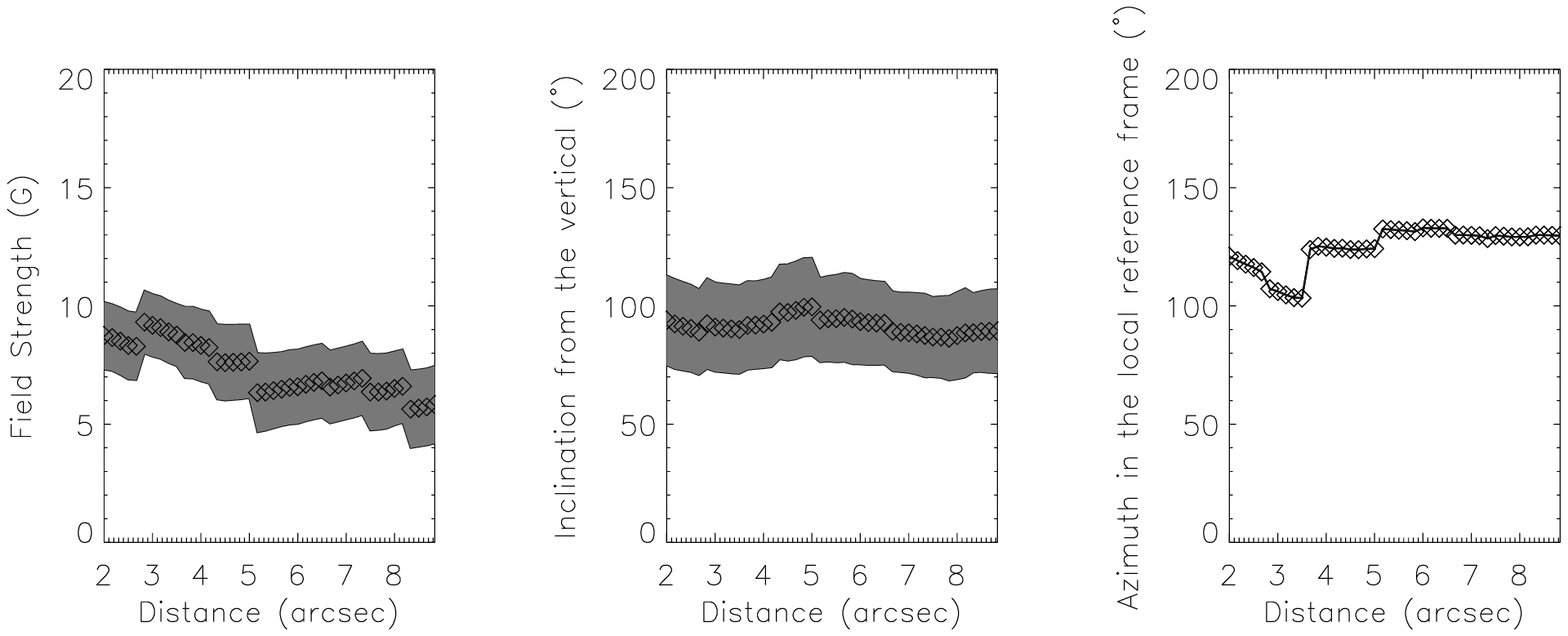}
\caption{
%Magnetic field strength, Inclination and Azimuth derived from the Stokes parameters obtained with THEMIS/MTR data.
%The grey regions are the uncertainties around the average value. The inclination to the vertical is usually between 100 and 150 degrees indicating a field that is mostly 
%horizontal and at an angle of about 30 degrees from the plane of the sky.
%along the radial line drawn in Figure \ref{Fig:themis2}
  From left to right, plots of the inferred magnetic field strength (G),
inclination (degrees, 90$^\circ$ being horizontal) and azimuth (degrees, 90$^\circ$ being
in the plane of the sky). They are plotted along a cut through the
brightest prominence column  (similar to the column indicated by slit (a) in the Fig.\ref{Fig:sot1} right panel). The
 diamonds are the actual data, while the shaded region is the 3-sigma
 smoothed confidence region of the results. The azimuth is subject to a
180$^\circ$ ambiguity that has been solved ad-hoc but that invalidates the
computation of the confidence limits.
\label{Fig:graphs}}
\end{figure*}

 Figure \ref{Fig:graphs}   presents the  variation of magnetic field strength, inclination and azimuth along the  brightest  column in Figure \ref{Fig:themis2}. 
 The field strength in the bright column  is around 7.5 Gauss and  horizontal.  The azimuth is close to 100$^\circ$. This means that the magnetic field vector is mainly in the plane of the sky. This confirms previous results \citep{bommier1998}.  Observations of prominence footpoints observed on the disk have also shown that the field lines are tangent to the photosphere  \citep{lopez2006}.  Linear force free field extrapolations show that prominence plasma is supported by shallow dips in magnetic field lines like in  the Kippenhahn-Schl\"{u}ter model \citep{aulanier1998,dudik2008}. The feet  or footpoint of a prominence would be piles  of  dips in horizontal field lines. The existence of a  dip in magnetic field lines to  represent a prominence thread  has been discussed accordingly to the value of the $\beta$ plasma  by \cite{heinzel1999}. Our observation  of prominence fits  closely  the side view of a footpoint of prominence (or barb)  modeled by dips of magnetic field lines  in  Figure 5 of   \cite{dudik2012}.

 The global shape of the foot is like an ``anvil'' and each portion of the field lines is  horizontal and roughly  in the plane of the sky.

\section{Fast Magnetosonic  Wave Model}

%Transverse oscillations have been detected in  vertical  bright regions or columns in the pillar   observed at  SAC Peak and with Hinode. We interpreted them as magneto sonic fast waves according to their  radial propagation  with respect to the magnetic field direction.     Dopplershifts indicate motions along the LOS.  The bright columns are pile-up of dips of field lines.

%In Ca II intensity  trains of waves are observed propagating upwards in the prominence with a projected phase speed of around $10~\mathrm{km~s^{-1}}$, a wavelength of $2000~\mathrm{km}$, and a period of $260~\mathrm{s}$. 

%This intensity variations in the CaII filters are interpreted as plasma compressions and  followed by rarefactions. Additionally, the propagation of the waves are perpendicular to the magnetic field. The observations show a train of waves propagating upwards in the prominence with a projected phase speed of $10~\mathrm{km~s^{-1}}$.

%%%%%%%%%%%%%%%%%%%%%%%%%%%%%%%%%%%%%%%%
\subsection{Theoretical Phase Velocity of Waves}

Propagating waves have been detected in three vertical bright columns observed by Hinode/SOT and the DST. In the column with the brightest \ion{Ca}{2} intensity (analyzed with slit a, Fig.~\ref{Fig:sot1}) the observed wave train propagates upwards in the prominence with a projected phase speed of around $10~\mathrm{km~s^{-1}}$, a projected wavelength of $2000~\mathrm{km}$, and a period of $277~\mathrm{s}$ as described in Sec.~\ref{Sect:WaveObs}. These intensity variations in the \ion{Ca}{2} filters are interpreted in terms of plasma compressions and rarefactions.  According to the magnetic field measurements the propagation of the waves are mainly perpendicular to the horizontal local magnetic field. Prominences have a complex fine structure that consists of thin threads \citep{Lin2005} that extend along the magnetic field. The bright columns consist of many piled-up threads. The wave propagates in a highly non-homogeneous medium; however, the measured phase speed seems independent of the position of the wavefront (see  Section 3.2). 

%\S\ref{sec:kalman}). 

\begin{figure}[!ht]
\hspace{-1cm}\includegraphics[width=10cm]{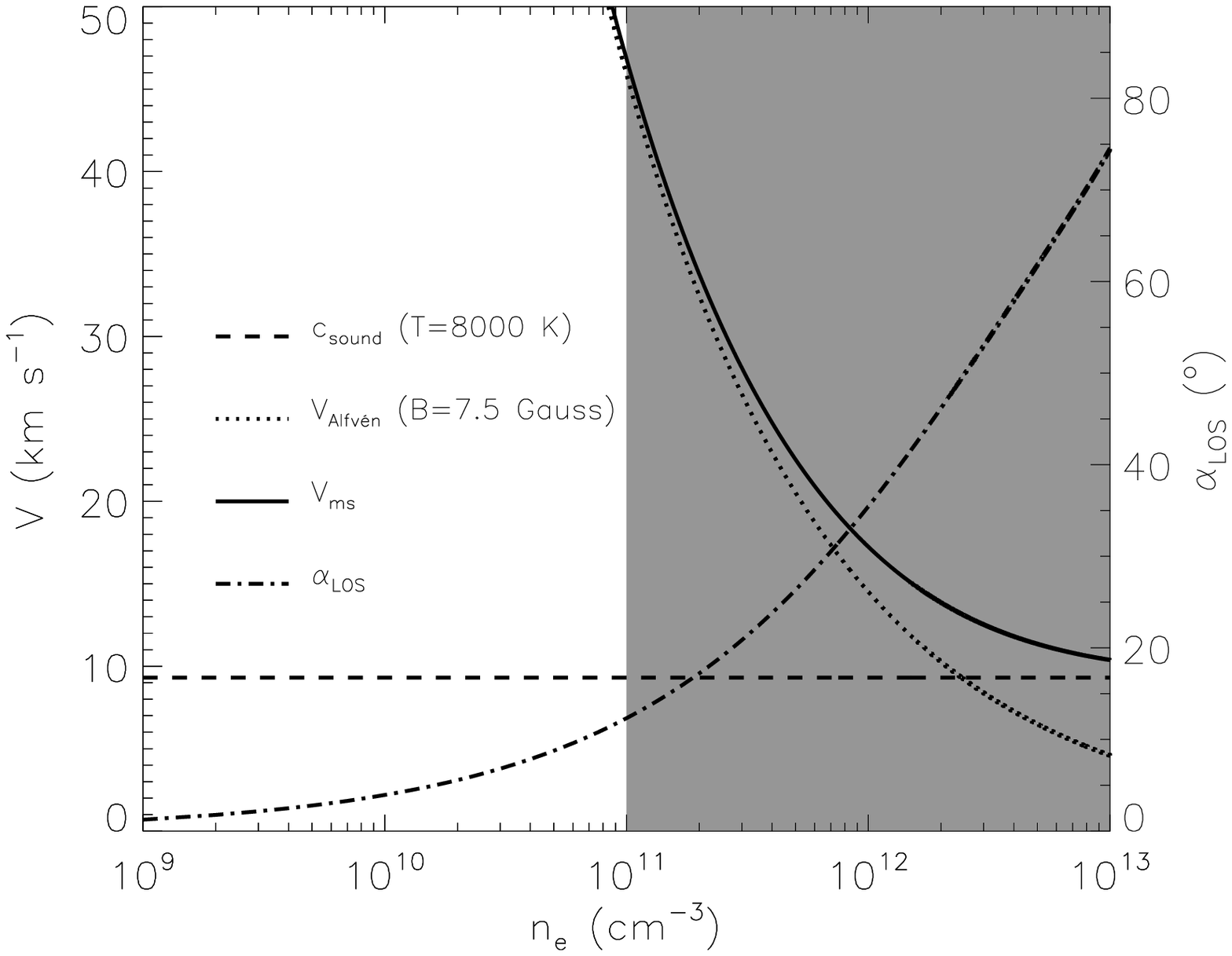}
\caption{Plot of the theoretical phase velocity of the magnetosonic mode (solid line), the slow mode (dashed line), the Alfv{\'e}n mode (dotted line) as a function of the electron number density of the prominence plasma. The magnetic field intensity used is $7.5~\mathrm{G}$ and a prominence plasma temperature of $8000~\mathrm{K}$. The typical prominence values of the electron number densities are $n_\mathrm{e} \sim 10^9 - 10^{11}~\mathrm{cm}^{-3}$ \citep[see,][]{Labrosse2010}. The dark area corresponds to larger values of $n_\mathrm{e}$ and we have plotted this values in order to have values of the theoretical magnetosonic speed comparable to the observed velocities. We have also plotted with a dot-dashed line the $\alpha_{LOS}=\arcsin (10/v_\mathrm{ms})$ defined as the angle should form the propagation direction of a magnetosonic wave with respect to the LOS in order to have a projected velocity of $10~\mathrm{km~s^{-1}}$ (the observed value). 
\label{Fig:luna}}
\end{figure}

As a first approximation we consider that waves propagate in a uniform medium, and we interpret the observed waves as fast magnetosonic modes. These wave modes propagate perpendicularly to the local magnetic field and the velocity perturbations are longitudinal, along the propagation direction. The restoring force of these waves are the magnetic and gas pressure, and it produces compressions and rarefactions of the plasma and the magnetic field intensity. Based on the observed properties, we explore the different wave modes that could be relevant to the situation by doing a theoretical phase velocity diagram. In Figure \ref{Fig:luna} we have plotted the sound speed, $c_\mathrm{sound}$, the Alfv{\'e}n speed, $v_\mathrm{Alfv\acute{e}n}$, and the magnetosonic speed given by 
\begin{equation}
\label{eq:magnetosonicspeed}
v_\mathrm{ms}=\sqrt{v_\mathrm{Alfv\acute{e}n}^2+c_\mathrm{sound}^2}~,
\end{equation}
as a function of the electron number density, $n_\mathrm{e}$ and the averaged magnetic field intensity of our observations of $7.5~\mathrm{Gauss}$. In the range of typical prominence values of electron number density $n_\mathrm{e} \sim 10^9 - 10^{11}~\mathrm{cm}^{-3}$ \citep{Labrosse2010} the plasma-$\beta$ is small and $ v_\mathrm{Alfv\acute{e}n} > c_\mathrm{sound} $. Our observations reveal a projected speed of the order of $10~\mathrm{km~s^{-1}}$ that is much smaller than the magnetosonic speeds. In fact, it is similar to $c_\mathrm{sound}$ corresponding to the slow modes. However, the slow modes propagates mainly along the magnetic field in the range of small $\beta$, that is perpendicular to the observed direction. The discrepancy between the observed and theoretical values could be explained by the projection effect. The waves form an angle with respect to the LOS, $\alpha_{LOS}$. In the same Figure \ref{Fig:luna} we have plotted the angle $\alpha_{LOS}$ necessary to have a projected $v_\mathrm{ms}$ velocity of $10~\mathrm{km s^{-1}}$ defined as $\alpha_{LOS}=\arcsin (10/v_\mathrm{ms})$. This indicates that for the range of typical prominence densities the wave moves in a direction forming an small angle with respect to the LOS of $\alpha_{LOS}< 15^\circ$ (or $\alpha_{LOS}> 165^\circ$). These angles are very extreme indicating that for typical prominence values the propagation is mainly along the LOS. Another possibility is that the prominence electron density is larger than  $10^{11}~\mathrm{cm}^{-3}$ and the projection angle is small. For example with $n_\mathrm{e}=5 \times10^{11}~\mathrm{cm}^{-3}$ the magnetosonic speed is of around $22~\mathrm{km s^{-1}}$ and $\alpha_{LOS}=27^\circ$. Thus, a combination of a relatively small $\alpha_{LOS}$ and relatively large prominence density can explain the small phase speed of the observed magnetosonic waves.

\subsection{Model of Waves}

The propagating waves seem to be confined in the cool prominence column. This could be because we only observe the cool plasma in our observations, and the visibility of the waves are difficult to observe outside the cool prominence. However, theoretical models predict that the waves are truly confined in the prominence, as described by \citet{joarder1992a,joarder1992b,joarder1993} and \citet{oliver1992,oliver1993}. These authors modeled the prominence as a slab of cool plasma using different configurations of the magnetic field. These systems exhibit many normal modes that are confined or trapped in the horizontal direction but propagate vertically. These works concentrated on the large wavelength limit, $k_z L \ll 1$, where $k_z$ is the vertical wavenumber and $L$ is the length of the prominence field lines. In that situation the whole prominence oscillates in phase in the vertical direction. However, in the waves described in this work the wavelength is short compared with the length of the prominence magnetic field so that $k_z L \gg 1$. A detailed normal mode analysis of the system is out of the scope of this paper;  it will be a subject of a future work.

\begin{figure}[!ht]
\hspace{-1cm}\includegraphics[width=10cm]{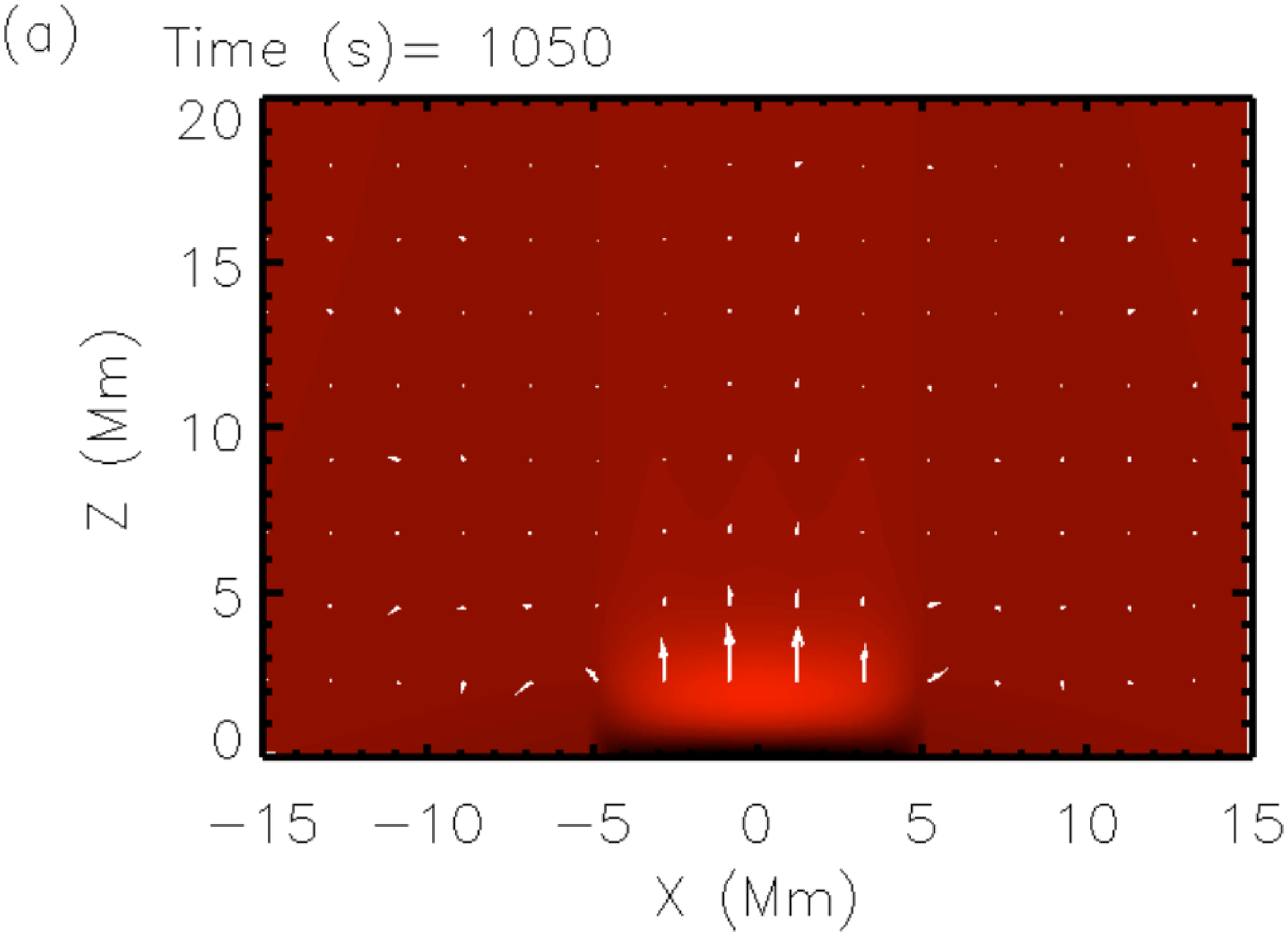}\\

\hspace{-1cm}\includegraphics[width=10cm]{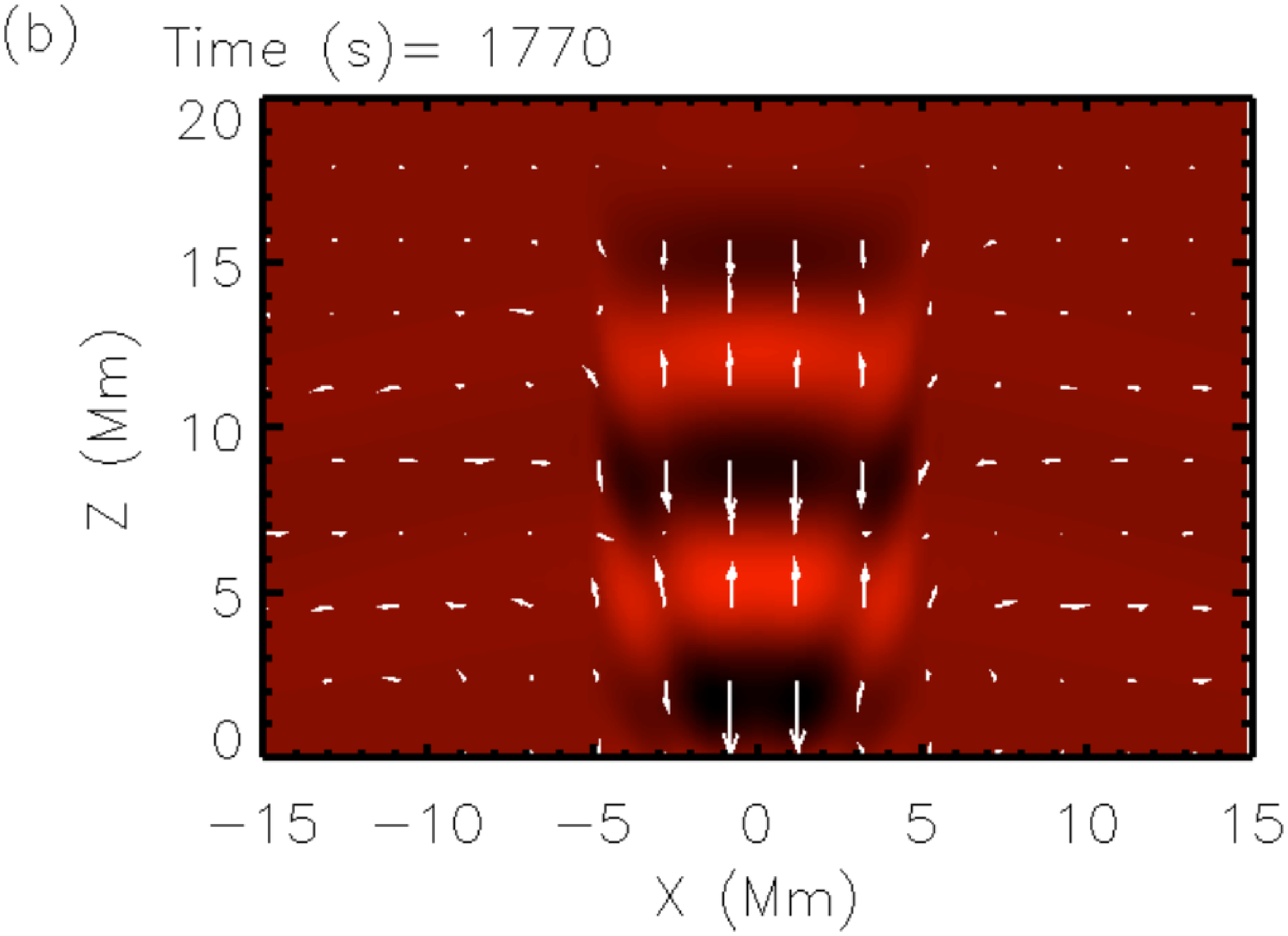}

\caption{Time-dependent simulation results of a wave train traveling along the prominence. The prominence slab is placed between $x=-5~\mathrm{Mm}$ and $x=5~\mathrm{Mm}$ with an uniform electron density of $10^{11}~\mathrm{cm}^{-1}$. The magnetic field is horizontal and uniform $\vec{B}=7.5~\hat{x}$. In (a) the wavefront appears from below and in (b) the full wave train has appeared traveling upwards. The gas pressure (color) shows almost planar wavefronts confined in the prominence column. Similarly, the velocity field (vectors) is almost vertical inside the prominence with the largest values, whereas is almost horizontal outside with smaller values. 
\label{Fig:luna2}}
\end{figure}

For the present work, we  have modeled the prominence as an uniform plasma slab with  a
uniform horizontal magnetic field (see Figs. \ref{Fig:luna2}a and \ref{Fig:luna2}b). The average magnetic field intensity is set to 7.5 Gauss. 
We have made a time dependent simulation of the system described by \citet{joarder1992b}. The model consists of a vertical prominence slab with a horizontal magnetic field transverse to the prominence slab. Gravity is neglected because it has little influence on the waves \citep{oliver1992}. In this simulation we try to understand how the waves propagate in the prominence structure and not how the waves are produced and reach the prominence. For this reason we have considered the prominence slab with an  infinite vertical extension and we have not modeled the photosphere, chromosphere, and transition region. The model atmosphere is uniform, the top of the photosphere is placed at $z = 0$. The driver perturbs the $v_z$-component of the velocity and consists in a planar pulse located at $z=-10~\mathrm{Mm}$ with a Gaussian shape, $v_z=e^{-(\frac{z+10}{3})^2}$. The width of the Gaussian is 3 Mm. We have checked several driver shapes and all produce similar results. The driver oscillates three times with a period of $300~\mathrm{s}$ that is similar to the observed one. The prominence-slab is placed between $x=-5~\mathrm{Mm}$ and $x=5~\mathrm{Mm}$, with a total width of $10~\mathrm{Mm}$. The electron density of the prominence is set to $n_\mathrm{e}=5\times10^{11}~\mathrm{cm}^{-3}$, $100$ times larger than that of the surrounding corona. The temperature of the prominence plasma is $8000~\mathrm{K}$ and the corona is $10^6~\mathrm{K}$. We fulfill the pressure imbalance condition $\rho_\mathrm{p} T_\mathrm{p} = \rho_\mathrm{c} T_\mathrm{c}$, where the p and c subscripts refer to prominence and coronal medium. The equilibrium magnetic field is uniform and horizontal, $\vec{B} = B_0 \hat{x}$, with $B_0=7.5~\mathrm{Gauss}$. The length of the field lines are 100 Mm that is in agreement with most of the theoretical modeling. We impose that the magnetic field perturbations are zero at both ends of the field lines representing the photospheric line-tying effect. The top and bottom boundaries of the simulation box are opened in order to mimic the infinite extension of the prominence in the vertical direction. We refer the reader to \citet{joarder1992b} for the equilibrium configuration and the equations describing the evolution.

The simulation shows that a train of waves traveling upwards confined in the prominence (see Figs. \ref{Fig:luna2}a and \ref{Fig:luna2}b). In Figure \ref{Fig:luna2}a, the wave appears from bottom. The gas pressure shows that this wave produces compressions and rarefactions of the prominence plasma. The velocity field oscillates vertically in the prominence plasma and horizontally in the corona, with the velocity in the prominence larger than that outside. The wave propagates vertically confined in the prominence body with a wave front almost planar. The motion of the plasma of the corona can be understood as follows: the coronal plasma reacts to the pressure changes of the prominence plasma. An increase of the plasma pressure in one selected  position forces  the coronal plasma to move away from this region along the magnetic field lines.  Similarly a decrease of the prominence plasma pressure forces  the coronal plasma to move towards that region. In Figure \ref{Fig:luna2}b, a three-wavelengths long wave train has appeared and travels upwards. In the rear of the train a small perturbation appears with a short wavelength. These are associated with an overtone also excited by the driver. The amplitude of these small perturbation is small in comparison with the upward propagating waves. The phase speed of the upwards waves is $22.6~\mathrm{km~s^{-1}}$, which coincides with the speed of the magnetosonic waves as we see in Figure \ref{Fig:luna}. We have performed several numerical experiments with different configurations and in all the cases the propagation speed coincides with the magnetosonic velocity. In \citet{joarder1992b,joarder1993}, the authors found that the short period oscillations are essentially trapped magnetosonic waves reflecting off the boundaries of the prominence slab and propagating along it. The normal mode excited in the simulations can be identified by the so-called string mode fIF. This mode is classified as internal mode by \citet{joarder1992b} where the velocity of the perturbation is mainly located inside the prominence. \citet{oliver1993} classified more accurately the normal modes of a prominence and found that the fIF modes are hybrid. This means that removing the coronal or prominence medium the mode remains, but with a different frequency and shape.

\begin{figure}[!ht]
\centering\includegraphics[width=9cm]{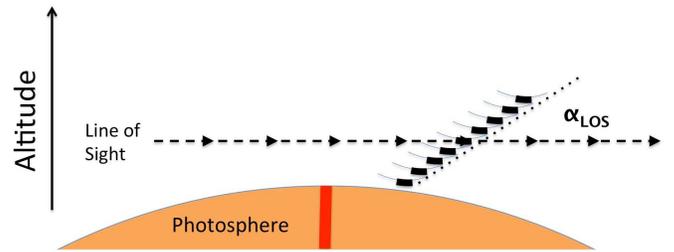}
\caption{Sketch of the possible configuration of the pillar and the observation. The pillar consists in threads located at dips that are pilled up. The pillar direction (dotted line) is inclined with respect to the local vertical to the solar surface (orange surface). The pillar is placed behind the solar limb (red line). The dashed line is the LOS with the observer placed at the left hand side of the plot. The LOS forms an angle $\alpha_\mathrm{LOS}$ with respect to the pillar. In the situation of a vertical pillar placed at the limb the angle is $90^\circ$. The inclination of the pillar and the position behind the limb make an angle, $\alpha_\mathrm{LOS}$ which is  smaller than $90^\circ$.
\label{Fig:kucera}}
\end{figure}

Our simulations demonstrate that the vertically propagating waves transverse to the magnetic field are magnetosonic waves ducted along the prominence column and the treatment of waves in an homogeneous medium is applicable. The observed prominence is quiescent and seems quite static in the observed time. This indicates that the plasma $\beta$ is small and then $ v_\mathrm{Alfv\acute{e}n} > c_\mathrm{sound} $ and the observed waves are fast magnetosonic modes. In order to have a projected magnetosonic speed comparable to the observed velocity the projection angle $\alpha_\mathrm{LOS}$ should be small. In fact, assuming that the prominence density have a typical $n_\mathrm{e}$, $\alpha_\mathrm{LOS} < 15^\circ$. As the wave is ducted in the prominence this indicates that the column of cool plasma forms and small angle with respect to the LOS. This could be associated to that the column is not placed exactly at the solar limb at the moment of the observations. In Figure \ref{Fig:EUVIA195} we see that the dark pillar is considerably behind the limb. It is almost 150 arcsecs behind, that is almost $10^\circ$ with respect to the Sun's center. This means that the local vertical to the solar surface at the pillar position is $80^\circ$ instead of $90^\circ$ with respect to the LOS. Additionally the bright column forms part of the prominence foot or barb, and such structures are often very elongated when seen on the disk but are not very tall when seen at the limb, indicating large inclinations of those structures with respect to the solar surface. In Figure \ref{Fig:EUVIA195} we can see that the horizontal extension of the pillar is of order of 50 arcsecs, and the vertical extension is less than 50 arcsecs from Figure \ref{Fig:sot1}. Then this means that the pillar is inclined more than $45^\circ$ with respect to the vertical direction. Numerical simulations indicates that prominence foot consist of many horizontal thread pilled to form inclined structures with respect to the solar vertical \citep[see Fig. 7 by][]{dudik2012}. In Figure \ref{Fig:kucera} we have included a sketch of the configuration of the pillar in  the observation. The prominence foot consists in a pillar of threads that are placed in the dips of the magnetic field lines. The pillar is behind the limb (red line) as the observations reveal (see Fig. \ref{Fig:EUVIA195}). The pillar has also an inclination with respect to the local vertical. Thus, it is plausible and consistent with our model that the oscillating columns form a small angle with respect to the LOS ($\approx 30^\circ$) and have a relatively large density  (a few times $10^{11}$~cm$^{-3}$).

%In a second pillar we observed projected velocities of up to $25~\mathrm{km~s^{-1}}$. This could indicate that the pillar has less inclination (contained in the plane of the sky) than the first bright pillar.

The DST observations reveals Doppler velocities (see \S \ref{sec:dst-observations}). These velocities are non-zero projections of the oscillation velocity along the LOS. This can be explained also by the inclination of the propagation of the waves with respect to the LOS. In the magnetosonic waves the oscillation and propagation direction are the same. Thus, in the case $\alpha_\mathrm{LOS} \neq 90^\circ$ the velocity of oscillation has a projection along the LOS.

Along the third slit we analyzed (slit  c) we observed downward motion, and  visually it appears that some waves move upwards and then reflect back downwards. These motions can be attributed to the propagation of the waves in an non-homogeneous medium. The study of the propagation of the waves in such a complicated medium will be a subject of a future work.

\section{Conclusion}

A quiescent prominence has been observed with Hinode (\ion{Ca}{2} H, H$\alpha$), Sac Peak (H$\beta$ and  H$\alpha$) over a time period of 4 hours,  THEMIS/MTR (vector magnetograms in He D$_3$), SDO/AIA  (193~\AA, 304~\AA), and STEREO-A/EUVI (195~\AA) on October 10 2012. The small field of view of the later instruments is centered on one large foot of the prominence. This foot  appears in 304 \AA, as a large quasi-vertical pillar with material flowing on each side along horizontal field lines. The polarimetry in the He D$_3$  line  obtained by the observations of THEMIS in the MTR mode allows us to derive the magnetic field strength, the inclination and the azimuth in the region of the prominence observed. The magnetic field is mainly horizontal with a field strength around 5 -10  Gauss.

The observed waves propagate perpendicularly to the magnetic field  
according to the measurements of the magnetic field by THEMIS. The  
observed phase speeds are below 10 km~s$^{-1}$, the periods are around 300 s,  
and the wavelengths are about 2000 km. Our simulations reveals that fast  
magnetosonic waves are ducted along the prominence foot moving  
upwards and probably away from the observer. These waves produce compressions and rarefactions of the  
plasma as the observations. The direction of the propagation are  
perpendicular to the magnetic field as expected. We conclude that the  
observed waves are fast magnetosonic waves ducted in the prominence  
foot. We can explain the small projected observed phase speed as a combination of a relatively small $\alpha_{LOS}$ and relatively large prominence density.

%{\bf Manuel: the field line of the feet are horizontal and form a small angle with LOS.. that is OK . Is that what you want to explain earlier??
%We conclude that the observed waves are fast magnetosonic waves ducted in the prominence feet. The prominence feet should form an small angle with respect to the LOS indicating an important inclination with respect to the solar vertical. 
%Additionally the density of the plasma column could be in the range of large electron number densities of the typical values or even larger, $n_\mathrm{e} \gtrsim 10^{11}$ but keeping the $\beta$ small.}

In some cases it seems that the wavefronts propagates downwards after a reflection at some height. This could be associated with non-homogeneities of the plasma and magnetic field. However, the waves are observed emanating from the direction of the solar surface. We cannot identify the driver of these waves. We speculate that they could be associated with rapid spicule jets, magnetic reconnection in the photosphere or just above, due to the emerging of small flux close to the prominence, or that the waves are associated with the characteristic 5-minutes photospheric oscillations tunneled to the chromosphere \citep{Heggland2011,Khomenko2013}. We observe waves during the entire 4 hour period of observation. This fact could be important for the coronal heating problem because these waves contributes to the AC heating. In a future work we will investigate the propagation of the waves in non-homogeneous plasmas and how different triggers can excite these waves in prominences.

%{\bf It is necessary to include more results as the Doppler information to complete this section.}

\acknowledgments

Hinode is a Japanese mission developed and launched by ISAS/JAXA, collaborating with NAOJ as a domestic partner, NASA and STFC (UK) as international partners. Scientific operation of the Hinode mission is conducted by the Hinode science team organized at ISAS/JAXA. This team mainly consists of scientists from institutes in the partner countries. Support for the post-launch operation is provided by JAXA and NAOJ (Japan), STFC (U.K.), NASA, ESA, and NSC (Norway). SDO data are courtesy of NASA/SDO and the AIA science team. 
The STEREO/SECCHI data are produced by an international consortium of the NRL, LMSAL and NASA GSFC (USA), RAL and Univ. Bham (UK), MPS (Germany), CSL (Belgium), IOTA and
IAS (France). The authors BS and AL would like to thank B.~Gelly the director of THEMIS and the technical team to have allowed to do these observations using the THEMIS instrument in Canaries Islands. TK would like to acknowledge support from the NASA Living with a Star program. ML gratefully acknowledge partial financial support by the Spanish Ministry of Economy through projects AYA2011-24808 and CSD2007-00050. This work contributes to the deliverables identified in FP7 European Research Council grant agreement 277829, " Magnetic connectivity through the Solar Partially Ionized Atmosphere", whose PI is E. Khomenko.

%
%
%=============================================================================================================
%\begin{thebibliography}{}

%\end{thebibliography}

%\bibliographystyle{apj}
%\bibstyle{apj}
%\bibdata{references}
%\bibliography{references}

\end{document}